\documentclass[lettersize,journal]{IEEEtran}
\usepackage[flushleft]{threeparttable}
\usepackage{algorithmic}
\usepackage{cite}
\usepackage{array}
\usepackage[caption=false,font=normalsize,labelfont=sf,textfont=sf]{subfig}
\usepackage{etoolbox}
\usepackage{algorithm}
\usepackage[]{footmisc}
\usepackage{textcomp}
\usepackage{enumitem}
\usepackage{stfloats}
\usepackage{url}
\usepackage{verbatim}
\usepackage{graphicx}
\usepackage{multirow}
\usepackage{mathtools}
\usepackage{float} % Add this to your preamble

\usepackage[flushleft]{threeparttable}
\usepackage{multicol}
\usepackage{array,colortbl,xcolor}
\usepackage{color,soul}
\usepackage{xcolor}
\usepackage{tikz}
\newcommand*\circled[1]{\tikz[baseline=(char.base)]{
		\node[shape=circle,draw,inner sep=0.4pt] (char) {#1};}}
\usepackage{mathtools}
\usepackage{amsmath,amsfonts}
\usepackage{amssymb,mathtools}

	\usepackage{colortbl}
	\usepackage{color,soul}
	\usepackage{xcolor}
	\usepackage{soul}
	
	\definecolor{violetThree}{rgb}{0.69, 0.29, 0.99}
	\definecolor{red}{rgb}{1, 0, 0}
	\definecolor{pink}{rgb}{1.0, 0.12, 0.23}
	\definecolor{orangeThree}{rgb}{1.0, 0.50, 0.005}
	\definecolor{pink}{rgb}{1.0, 0.08, 0.58}
	
\begin{document}
		\title{{COBALT: Column-swapping Optimized Bit-serial Accelerator for LSTM Tasks}}
		\author{Mohd~Tasleem~Khan,~\IEEEmembership{Senior Member,~IEEE} 
			%\author{Mohd~Tasleem~Khan 

				%Corresponding author: Mohd Tasleem Khan ({\it{e-mail:}} M.T.Khan@hw.ac.uk).
				%Manuscript received on 31 December 2024;
				%}
		}
		\maketitle
		
		\begin{abstract}
			
			Long Short-Term Memory (LSTM) networks continue to be widely deployed for real-time sequential tasks on
			edge devices---{yet their computational demands challenge deployment on resource-constrained
				hardware.} This work introduces COBALT, a bit-serial compressed LSTM accelerator built {on a matrix
				(input)--vector (weight) reformulation of the standard circulant matrix-vector multiplication (MVM) with
				offset-binary coding.} A novel column-swapping scheme operates on partial products at the bit level when
			generated in pairs, systematically exposing redundancy across output rows to reduce the number of PP
			generators and selectors. {This redundancy is further exploited using a lightweight correction unit
				that derives a row's output directly from its paired row.} Additionally, for block-circulant MVMs,
			relocating the shift-accumulate and correction units of each sub-MVM further reduces resource usage.
			{The compressed network achieves weight compression of up to 93.6\% while
				maintaining accuracy on the TIMIT and LibriSpeech-100h benchmarks.} On a field-programmable gate array,
			COBALT achieves superior overall efficiency relative to state-of-the-art LSTM accelerators.
		\end{abstract}
		

		\section{Introduction}
		\IEEEPARstart{T}{he} {growing deployment of machine learning on embedded and Internet-of-Things (IoT) platforms has exposed a persistent mismatch between the computational demands of neural network (NN) inference and the strict budgets (e.g., power and/or latency) of edge hardware \cite{khan2026next}.} {Meeting real-time inference targets under these constraints typically requires dedicated, application-specific hardware acceleration rather than general-purpose processing \cite{nurvitadhi2016accelerating}.}
		
		Recurrent NNs (RNNs) are widely used for real-time sequential learning tasks, such as speech recognition and machine translation, due to their ability to model temporal dependencies \cite{yang2020survey}. Standard RNNs, however, suffer from the vanishing gradient problem, which limits their effectiveness on long sequences. Long Short-Term Memory (LSTM) networks address this through gating mechanisms that capture long-range dependencies \cite{hochreiter1997long}. {Transformer-based architectures now dominate large-scale sequence modeling, often outperforming
			LSTMs when compute and memory are abundant \cite{tay2022efficient}. On resource-constrained edge and IoT
			devices, however, LSTMs remain preferred for streaming, latency-critical inference: their constant
			per-step memory footprint and linear complexity with sequence length avoid the growing cost that
			quadratic-complexity attention incurs under tight power and latency budgets
			\cite{ansari2025beyond,tay2022efficient, khan2026next}. This is particularly acute on
			field-programmable gate arrays (FPGAs) and microcontrollers} {with only kilobytes-to-low-megabytes
			of on-chip memory, where a growing key-value cache is often undeployable, while an LSTM's hidden and
			cell state stay fixed regardless of sequence length.} {This continued relevance is reflected in
			recent RNN hardware accelerators and behavioral models for real-time edge and RF applications
			\cite{gao2022spartus, zhang2023block, khan2026next}, and in TinyML deployments of LSTM-based models on
			microcontroller-class hardware for time-series and sensor tasks \cite{tinyml_lstm_soilmoisture}. As
			mobile and IoT devices increasingly demand low-latency, always-on inference, exclusive reliance on cloud
			processing is unsustainable,} {motivating the shift of LSTM computation onto resource-limited
			devices.}
		
		LSTM networks have been widely implemented on specialized hardware to enable efficient edge computation. FPGAs are particularly well suited for LSTM inference due to their inherent parallelism and efficient use of on-chip memory and DSPs \cite{gao2018deltarnn, wang2018c, li2019rnn, cao2019efficient, wang2019lstm, gao2022spartus, li2023fpga, alhartomi2023low, kim2023v, kim2024auto, sun2021fpga}. Nevertheless, LSTM inference remains challenging due to large parameter counts and several matrix--vector multiplications (MVMs) and element-wise multiplications (EWMs), which place high demands on memory and computation. Large models often exceed on-chip capacity, requiring external DRAM access that increases power consumption \cite{han2016eie}, whereas on-chip SRAM offers faster, more energy-efficient storage \cite{wang2017accelerating}. To alleviate these memory constraints, compression techniques such as parameter sharing \cite{han2016eie}, block-circulant weight matrices with hardware co-design \cite{li2019rnn}, and pruning \cite{wang2019lstm, kim2023v} have been explored, {though often at the cost of some accuracy relative to unconstrained models \cite{ravanelli2019pytorch}.} For instance, Spartus \cite{gao2022spartus} leverages spatio-temporal sparsity for ultra-low-latency inference, while structured top-$k$ pruning \cite{wang2019lstm} and Viterbi-based pruning \cite{kim2023v} further reduce memory usage on embedded devices.
		
		Although compression techniques reduce the memory footprint of deep NNs, structured parameterizations
		based on low-displacement-rank matrices—particularly circulant and block-circulant matrices—offer an
		effective alternative for parameter-efficient learning \cite{wang2017accelerating}. Their algebraic
		structure eliminates weight redundancy, cuts storage complexity from quadratic to linear, and enables
		FFT-based matrix operations \cite{li2018efficient}. {Block-circulant adapters recently achieved
			competitive parameter-efficient fine-tuning for large language models with far fewer parameters and
			operations than low-rank adaptation} \cite{chen2025correlating, ding2025block, ding2025parameter}. Larger
		block sizes trade accuracy for compression, making block-wise designs a practical
		efficiency--performance balance well suited to scalable FPGA-based NN accelerators.
		
		%Although compression techniques mitigate memory demands, low-displacement rank matrices—such as circulant and block-circulant forms—provide a structured alternative by removing weight redundancy and reducing storage \cite{wang2017accelerating, li2018efficient, yue2020sticker}. They support efficient fine-tuning in large language models via compact Fourier-based updates \cite{chen2025correlating, ding2025block}. Circulant operations are especially efficient when model sizes are powers of two \cite{li2018efficient}. A trade-off exists: larger blocks yield higher compression but reduce accuracy \cite{alhartomi2023low}. These properties make circulant matrices well-suited for scalable, structured FPGA accelerators \cite{li2018efficient}. 

		In LSTMs, computational complexity is dominated by resource-heavy MVMs and EWMs, both of which rely on bulky multipliers. Prior work has explored circulant weight matrices to reduce memory accesses and optimize MVMs via time-multiplexing \cite{wang2017accelerating}. Multiplier complexity has been further reduced through multiplierless optimizations based on distributed arithmetic (DA) in bit-serial form \cite{khan2022architectural,alhartomi2023low,khan2024digit}. DA computes an inner product (IP) bit-serially: at each cycle, a bit-slice of one vector addresses a look-up table (LUT) pre-loaded with the partial-product (PP) sums of the other vector, and the LUT output is accumulated by a shift-accumulate (SA) unit, which combines the PP sums over a number of clock cycles. It applies either two's complement (TC), which treats the sign bit-slice separately via a conditional adder/subtractor in the SA unit, {or offset-binary coding (OBC), sign handling is folded into a single constant offset loaded once at the start of accumulation, allowing every bit slice to use uniform addition and halving the LUT size. Unlike TC, OBC's PPs are symmetric, enabling more efficient implementation.}
		
		{Hardware LUTs with PP generators (PPGs) and selectors (PPSs) enable efficient LSTM accelerators
			\cite{khan2022architectural,alhartomi2023low,khan2024digit}. In \cite{alhartomi2023low}, TC-DA LUTs generate full or partial parallel PPs via a separate PPGs across all four LSTM gates, causing adder and register counts to grow exponentially with model size. Its DA-based MVM formulation is confined to circulant weight matrices, with PPGs, PPSs, and adders instantiated independently for each row pair and one SA unit per sub-MVM. To curb this growth, two pipelined high-radix OBC-DA serial-LUT architectures were proposed \cite{khan2022architectural,khan2024digit}, but at the cost of additional clock-cycle latency in tapped-delay units that likewise scale with model size. Across these designs, pipelining and unfolding trade hardware complexity for latency, while registers persist throughout the circular shifters, tapped-delay units, and LUT pipelines, with no specific optimization for block-circulant structures.}
		
		{Existing efforts have reduced MVM and EWM cost using circulant weight matrices, employing them as
			sub-MVMs for block-circulant MVMs. However, none has reformulated this multiplication into an equivalent
			input-matrix--weight-vector product, which reduces the number of distinct PPGs required to realize LSTM
			networks with minimal hardware cost. To address this gap, this work adopts the input-matrix--weight-vector
			product, structurally eliminating the hardware overhead common to these designs.} {Building on this
			approach, COBALT is proposed as a highly efficient bit-serial LSTM accelerator for compressed-matrix
			inference,} with key contributions as:

		\begin{itemize}
			\item The weight–input MVM is reformulated as an input–weight product and analyzed using the OBC scheme for training.
			\item PPs are generated from input pairs using PPGs and selected through PPSs, achieving the lowest cost.
			\item {Pairwise column-swapping of PPs} exposes redundancy across output rows, reducing the PPGs, PPSs, and adders needed by sharing them across row pairs.
			\item {A lightweight correction scheme that reuses the paired row's PPs sum, eliminating the need for a dedicated adder when generating redundant rows.}
			\item Relocating the SA unit of each sub-MVM in block-circulant MVMs further reduces adders and registers.
		\end{itemize}

		%The remainder of the paper is organized as follows. Section II introduces the COBALT and matrix transformation with OBC analysis. Section III presents the fixed-point training methodology and architectural details. Section IV describes the proposed LUT optimization and architecture. Section V compares the performance of various LSTM accelerators. Finally, Section VI concludes the paper.
		
		{The remainder of this paper is organized as follows. Section~II briefly introduces COBALT and the
			circulant and block-circulant MVM reformulation, followed by the OBC analysis. Section~III presents the
			hardware optimization and proposed accelerator for the compressed LSTM. Section~IV presents the results
			and discussion, and Section~V concludes the paper.}

		\section{LSTM Reformulation, OBC Analysis {and Training Strategy}}\label{SectionI}
		At time-step $t \in (1,2,...,T)$,  the following equations capture the operation of a standard LSTM layer \cite{hochreiter1997long}:
		\begin{align}
			\label{eq:lstm1}
			\textit{\textbf{s}}^t &= {\phi}(\mathbf{W}_s\mathbf{x}^t + \mathbf{R}_s\mathbf{y}^{t-1} + \mathbf{b}_s) &  (s=i,f,o,z)\\
			\mathbf{c}^t &= \mathbf{z}^t\odot\mathbf{i}^t + \mathbf{c}^{t-1}\odot\mathbf{f}^t & {\textnormal{memory cell}}\\
			\mathbf{y}^t &= {\textnormal{tanh}}(\mathbf{c}^t)\odot\mathbf{o}^t & {\textnormal{hidden state}}
		\end{align}
		\noindent 
		Let ${\bf{W}}_s$ and ${\bf{R}}_s$ represent the parameter matrices of dimensions $N_2 \times N_1$ and $N_2 \times N_2$, corresponding to the input weights and recurrent connections, respectively. The input vector ${\bf{x}}^t$ and the recurrent vector ${\bf{y}}^{t-1}$ are of size $N_1 \times 1$ and $N_2 \times 1$ respectively. The vector ${\mathbf{b}_s}$ denotes the bias terms. The non-linear activation function $\phi(\cdot)$ is sigmoid $\sigma(\cdot)$ for the forget (${\bf f}^{t}$), input (${\bf i}^{t}$), and output (${\bf o}^{t}$) gates, while the block gate (${\bf z}^{t}$) uses the hyperbolic tangent tanh$(\cdot)$. For any input $x$, $\sigma(\cdot)$ and tanh$(\cdot)$ are, respectively, defined as 
		\begin{equation}\label{eq4}
			\begin{aligned}
				\sigma(x) & = \frac{1}{1 + e^{-x}} {\in [0, 1],} \hspace{0.1cm} \text{and} \\
				\textnormal{tanh}(x)  &  = \frac{1 - e^{-x}}{1 + e^{-x}} {\in [-1, 1]}
			\end{aligned}
		\end{equation}
		\begin{figure}[t]
			\centering
			\includegraphics[width=0.95\linewidth]{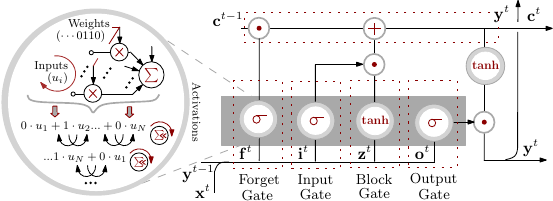}
			\caption{{Concept of the COBALT with a standard LSTM layer.}}\label{fig1}
		\end{figure}
		
		\noindent Based on (1), COBALT is motivated by the high computational cost of the MVMs, $\mathbf{W}_s\mathbf{x}^t$ and $\mathbf{R}_s\mathbf{y}^{t-1}$, 
		{which dominate the execution time and energy consumption of a standard LSTM. COBALT exploits bit-level redundancy in the weight matrices to reformulate MVMs as bit-serial operations, reducing computational complexity.} Fig. 1 provides an overview of the proposed 
		COBALT architecture.
		\subsection{MVM Reformulation with Circulant Matrices}
		For given $N_1$ and $N_2$ dimensions, the total number of parameters required for an LSTM layer, as per (1), is $4(N_1N_2 + N_2^2 + N_2)$. When $N_1 = N_2 = N$, this simplifies to $8N^2 + 4N$. {Circulant matrices efficiently reduce LSTM memory requirements \cite{wang2017accelerating}. Here, the weight matrices in the MVMs of (1) are replaced with circulant matrices, except for the block gate ($s\neq z$), which is more sensitive to numerical errors \cite{wang2017accelerating}. This replacement reduces the number of parameters to $2N^{2} + 6N$, while the biases remain at $4N$, resulting in a total of $2N^{2} + 10N$ parameters. Overall, this achieves a reduction of approximately $4\times$ in the number of parameters. However, this change requires additional steps, since circulant matrices are typically square while LSTM parameter matrices are not necessarily so. The structure typically depends on the ratio of $N_1/N_2$ (or $N_2/N_1$). When this ratio is an integer multiple, the matrix can be decomposed into sub-circulant blocks \cite{li2018efficient}. If the dimensions differ slightly (i.e., the ratio is close to 1), the input vector $\mathbf{x}^t$ can be padded or truncated with minimal impact on accuracy, owing to the resilience of structured representations \cite{wang2017accelerating}. Thus, a normal square matrix ${\bf W} = [w_{mn}]$, where $1 \leq m, n \leq N$, can be transformed by a circulant matrix generated from its first column, known as the primitive vector ${\bf w}_1 = \{w_m\}_{m=1}^{N}$.} The remaining columns of ${\bf{W}}$ are generated by cyclically shifting the elements of ${\bf{w}}_1$ to the right. Thus, each column vector ${\bf{w}}_n$ of ${\bf{W}}$ can be expressed as ${\bf{w}}_n = [w_{1n}, w_{2n}, \dots, w_{Nn}]^T$, where $mn = (m-n)\textnormal{mod}(N) + 1$. Consequently, the MVMs in (1), can be formulated as 
		\begin{equation}\label{eq7}
			{{\bf{v}}}={\bf{W}}{\bf{u}} \Rightarrow
			\begin{bmatrix}
				{{v}}_{1}\\
				{{v}}_{2}\\
				{{v}}_{3}\\
				\vdots\\
				{{v}}_{N}\\
			\end{bmatrix}=
			\begin{bmatrix}
				{{w}}_{1} & {{w}}_{N}  & \dots & {{w}}_{2} \\
				{{w}}_{2} & {{w}}_{1}  & \dots & {{w}}_{3} \\
				{{w}}_{3} & {{w}}_{2}  & \dots & {{w}}_{4} \\
				\vdots & \vdots  & \ddots & \vdots \\
				{{w}}_{N} & {{w}}_{N-1}  & \dots & {{w}}_{1} \\
			\end{bmatrix}
			\begin{bmatrix}
				{{u}}_{1}\\
				{{u}}_{2}\\
				{{u}}_{3}\\
				\vdots\\
				{{u}}_{N}\\
			\end{bmatrix}
		\end{equation}
		where ${\bf{W}}$ is the $N \times N$ circulant matrix, ${\bf{u}}=\{u_n\}_{n=1}^N$ and ${\bf{v}}=\{v_m\}_{m=1}^{N}$ are the input and output vectors, respectively. {Similar to (5), the circulant MVM can be reformulated by redistributing the elements of the input vector across all rows such that the first row of the parameter matrix remains fixed. In this reformulation, the circulant structure is assigned to the input matrix rather than the weight matrix, transforming the weight matrix ${\bf{W}}$ into a weight vector ${\bf{w}}$ and the input vector ${\bf{u}}$ into an input matrix ${\bf{U}}$, expressed as} 
		\begin{equation}\label{eq7n}
			\hspace{-0.2cm}{{\bf{v}}}={\bf{U}}{\bf{w}} \Rightarrow
			\begin{bmatrix}
				{{v}}_{1}\\
				{{v}}_{2}\\
				{{v}}_{3}\\
				\vdots\\
				{{v}}_{N}\\
			\end{bmatrix}=
			\begin{bmatrix}
				{{u}}_{1} & {{u}}_{2}  & \dots & {{u}}_{N} \\
				{{u}}_{2} & {{u}}_{3}  & \dots & {{u}}_{1} \\
				{{u}}_{3} & {{u}}_{4}  & \dots & {{u}}_{2} \\
				\vdots & \vdots  & \ddots & \vdots \\
				{{u}}_{N} & {{u}}_{1}  & \dots & {{u}}_{N-1} \\
			\end{bmatrix}
			\begin{bmatrix}
				{{w}}_{1}\\
				{{w}}_{N}\\
				{{w}}_{N-1}\\
				\vdots\\
				{{w}}_{2}\\
			\end{bmatrix}
		\end{equation}
		where the elements of $\mathbf{U}$ are given by $u_{mn}$ with subscript is expressed as $mn = (m + n - 2) \bmod (N) + 1$, and the elements of $\mathbf{w}$ are $w_{r_n}$ with subscript is given by, ${r_n} = (1 - n) \bmod (N) + 1$. Unlike ${\bf W}$, ${\bf U}$ is constructed by left cyclic shifts of a primitive vector ${\bf u} = \{u_n\}_{n=1}^N$, making each row a circular shift of the previous one. Re-writing (\ref{eq7n}) as IPs results in   
		\begin{equation}\label{eq9}
			v_{m}=\sum\nolimits_{n=1}^{N}u_{mn}w_{r_n},
		\end{equation}
		which leads to the following transformation of (1):
		\begin{align}\label{eq8}
			\textit{\textbf{s}}^t &= \phi(\mathbf{X}^t\mathbf{w}_s + \mathbf{Y}^{t-1}\mathbf{r}_s + \mathbf{b}_s)
		\end{align}
		{This shifts the computational load to just two circulant input and recurrent matrices, instead of eight separate weight matrices as in (1), enabling more efficient inference optimization, as discussed in the next Section.}
		
		For large $N$, an $N \times N$ circulant matrix can be partitioned into $p^2$ smaller $q \times q$ circulant matrices, where $q = N/p$, using a block circulant structure \cite{li2018efficient}. This ensures that the smaller circulant sub-matrices are evenly distributed across all rows and columns. Consequently, primitive vectors $p$ are generated by splitting the primitive input vector ${\bf{u}}$ into segments, denoted ${\bf{u}}^1, {\bf{u}}^2, \dots, {\bf{u}}^p$, where ${\bf u}^p = \{ {u_{l^{p}_n}} \}_{n=1}^q$, with the subscript given by ${l^{p}_n = (p-1)q + n}$. Each sub-circulant matrix ${\bf U}^p$ can be expressed in terms of ${\bf u}^p$ elements as ${\bf U}^p = \{ {u_{l_{mn}^p}} \}_{m,n=1}^{q}$, where the subscript {$l_{mn}^p$} is defined as
		\[
		{l^{p}_{mn}=(p-1)q + (m+n-2) \bmod (q) + 1}
		\]
		and $(m, n) \in \{1,2,...,q\}$ refers to the (row, column) indices of ${\bf{U}}^p$ respectively. Similarly, the weight vector ${\bf{w}}$ is split into $p$ smaller vectors, ${\bf{w}}^{1}, {\bf{w}}^{p}, \dots, {\bf{w}}^{2}$, where $\mathbf{w}^p = \left\{ {w_{r_{n}^{p}}} \right\}_{n=1}^{q}$, with subscript ${{r_{n}^{p}}}$ is defined as
		\[
		{r_n^{p} =  \bigl(1 -  (p-1)q - n \bigr) \bmod(q) + 1.}
		\]
		Likewise, the output vector ${\bf{v}}$ is partitioned into $p$ smaller vectors, ${\bf{v}}^{1}, {\bf{v}}^{2}, \dots, {\bf{v}}^{p}$, where ${\bf{v}}^p = \{ {v_{l^{p}_{m}}} \}_{m=1}^q$ with subscript given by {$l^{p}_{m}=(p-1)q+m$}. Thus, the block-circulant MVM is given by
		\begin{equation}
			\begin{bmatrix}
				{\bf{v}}^{1}\\
				{\bf{v}}^{2}\\
				{\bf{v}}^{3}\\  
				\vdots\\
				{\bf{v}}^{p}\\
			\end{bmatrix}=
			\begin{bmatrix}
				{\bf{U}}^{1} & {\bf{\Psi}}{\bf{U}}^{2} & {\bf{\Psi}}{\bf{U}}^{3}  & \dots & {\bf{\Psi}}{\bf{U}}^{p} \\
				{\bf{U}}^{2} & {\bf{U}}^{3} & {\bf{\Psi}}{\bf{U}}^{4}  & \dots & {\bf{\Psi}}{\bf{U}}^{1} \\
				{\bf{U}}^{3} & {\bf{U}}^{4} & {\bf{U}}^{5}  & \dots & {\bf{\Psi}}{\bf{U}}^{2} \\
				\vdots & \vdots & \vdots  & \ddots & \vdots \\
				{\bf{U}}^{p} & {\bf{U}}^{1} & {\bf{U}}^{2}  & \dots & {\bf{U}}^{p-1} \\
			\end{bmatrix}
			\begin{bmatrix}
				{\bf{w}}^{1}\\
				{\bf{w}}^{p}\\
				{\bf{w}}^{p-1}\\
				\vdots\\
				{\bf{w}}^{2}\\
			\end{bmatrix}.
			\label{eq:block_circ}
		\end{equation}
		Here, ${\bf{\Psi}}$ is the circular-shifter matrix with elements either $0$ or $1$. Its elements are defined by $\psi_{mn}=\delta[(m-n)\textnormal{mod}(q)-1]$, where $\delta[\cdot]$ is the Kronecker delta function. The purpose of ${\bf{\Psi}}$ is to rotate the elements of ${\bf{U}}^{p}$. Like block-circulant weight matrices \cite{wang2017accelerating}, ${\bf{\Psi}}$ in block-circulant input matrices appears in anti-lower triangular. It is evident from (\ref{eq:block_circ}) that the computation of each output vector ${\bf{v}}^{p}$ involves $p$ weight  vectors ${\bf{w}}^{p}$ and $p$ input sub-matrices ${\bf{{U}}}^{p}$. Thus, each ${\bf{v}}^{p}$ can be calculated sequentially over $p$ clock cycles, as per 
		
		\begin{align}
			\mathbf{v}^1 
			&= \mathbf{U}^1 \mathbf{w}^1 
			+ \Psi \mathbf{U}^2 \mathbf{w}^p 		
			+ \dots  
			+ \Psi^{p-1} \mathbf{U}^{p} \mathbf{w}^{2} \nonumber \\
			&= \sum\nolimits_{k=1}^{p} 
			\Psi^{\,k-1}\,\mathbf{U}^{k}\,
			\mathbf{w}^{\left((1-k+p)\bmod p\right)+1},
			\hspace{0.1cm} (\Psi^{0}=\mathbf{I}).
		\end{align}
		As per~(10), the sub-MVM operation for any \( p \) can be computed in parallel, thereby allowing~(\ref{eq:block_circ}) to be completed in \( p \) clock cycles. In scalar form, (10) can be expressed as
		\begin{equation}\label{eq:block} 
			v_{{l^i_m}} = \sum\nolimits_{j=1}^{p} \sum\nolimits_{n=1}^{q} u_{{l_{mn}^{(i-j)}}}w_{{r_n^j}}=\sum\nolimits_{j=1}^{p}v^{ij}_m
		\end{equation}
		where the subscripts {$l^i_m$, $r_n^j$} and {$l_{mn}^{(i-j)}$} are defined as
		\[
		{l^i_m=(i-1)q+m, \hspace{0.1cm} r_n^j = \big(1 - (j - 1)q - n \big) \bmod (q) + 1,}
		\]
		\[
		{l_{mn}^{(i-j)} = q ( (i - j) \bmod (p) ) + (m + n - 2) \bmod (q)  + 1}
		\] 
		respectively, and $(i, j) \in \{1,2,...,p\}$ refers to the (row, column) indices of ${\bf{U}}$ respectively. 
		Thus, ${{v}}^{ij}_{{m}}$ is defined as
		\begin{equation}\label{eq:ipc}
			{{v}}^{ij}_{{m}}=\sum\nolimits_{n=1}^{q}u_{{l_{mn}^{(i-j)}}}w_{{r_n^j}}
		\end{equation}
		\subsection{OBC-DA Analysis}
		{The OBC equivalent of (\ref{eq:ipc}) is derived starting from the standard TC form of the input as follows.} Any $B$-bit input $u_{{l_{mn}^{(i-j)}}}$ in TC form can be expressed as
		\begin{equation}\label{eq14}
			u_{{l_{mn}^{(i-j)}}}=-u_{{l_{mn}^{(i-j)},B-1}}+\sum\nolimits_{k=0}^{B-2}u_{{l_{mn}^{(i-j)},k}}2^{-(B-1-k)}
		\end{equation}
		where $u_{{l_{mn}^{(i-j)},k}}$ is the $k{\textnormal{th}}$-bit of $u_{{l_{mn}^{(i-j)}}}$, with $k=0$ denoting the LSB and $k=B-1$ the sign bit (MSB). Using OBC, each magnitude bit can be represented as $u_{{l_{mn}^{(i-j)},k}}=\frac{1}{2}\big[1+\Delta u_{{l_{mn}^{(i-j)},k}}\big]$ for $k=0,\ldots,B-2$, and the sign bit as $u_{{l_{mn}^{(i-j)},B-1}}=\frac{1}{2}\big[1-\Delta u_{{l_{mn}^{(i-j)},B-1}}\big]$, leading (\ref{eq14}) to
		\begin{equation}\label{eq14a}
			u_{{l_{mn}^{(i-j)}}}= \frac{1}{2}\bigg[\sum\nolimits_{k=0}^{B-1}\Delta u_{{l_{mn}^{(i-j)},k}}2^{-(B-1-k)}-2^{-(B-1)}\bigg]
		\end{equation}
		where 
		\begin{equation}\label{eq14b}
			\Delta u_{{l_{mn}^{(i-j)},k}}=(-1)^{\lfloor k/(B-1) \rfloor}\left (u_{{l_{mn}^{(i-j)},k}}-{\overline{u}}_{{l_{mn}^{(i-j)},k}}\right)    
		\end{equation}
		{where ${\overline{u}}_{l_{mn}^{(i-j)},k}$ is the ones complement of ${{u}}_{l_{mn}^{(i-j)},k}$ and $\lfloor \cdot \rfloor$ is the floor function.} Substituting (\ref{eq14b}) and (\ref{eq14a}) into (\ref{eq:ipc}) and simplifying the terms, we obtain
		\begin{equation}\label{eq14c}
			{{v}}^{ij}_{{m}}=\sum_{k=0}^{B-1}\sum_{n=1}^{q}\bigg(\frac{1}{2}w_{{r_n^j}}\Delta u_{{l_{mn}^{(i-j)},k}}\bigg)2^{-B_k}-\bigg(\sum_{n=1}^{q}\frac{1}{2}w_{{r_n^j}}\bigg)2^{-B_0}
		\end{equation}
		where $B_k=B-1-k$. Define
		\begin{equation}\label{eq14dd}
			{{d}}^{ij}_{m,k}=\sum_{n=1}^{q}\bigg(\frac{1}{2}w_{{r_n^j}}\Delta u_{{l_{mn}^{(i-j)},k}}\bigg), \hspace{0.2cm} {{d}}_{ot}=\sum_{n=1}^{q}\frac{1}{2}w_{{r_n^j}}
		\end{equation}
		Note that although $r_n^j$ is notated with superscript $j$ for consistency with (\ref{eq:ipc}), it is in fact independent of $j$: since $(j-1)q\bmod q=0$, $r_n^j=(1-n)\bmod q+1$ for every $j$. Consequently, $d_{ot}$ is a constant offset, independent of $j$.
		Finally, (\ref{eq14c}) simplifies to
		\begin{equation}\label{eq14d}
			{{v}}^{ij}_{m}=\sum\nolimits_{k=0}^{B-1}{{d}}^{ij}_{m,k}2^{-(B-1-k)}-d_{ot}2^{-(B-1)}       
		\end{equation}
		\begin{figure}[t]
			\centering
			\includegraphics[width=0.88\linewidth]{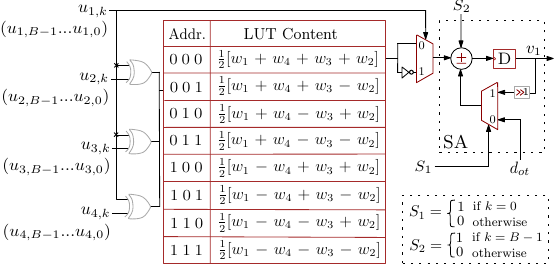}
			\caption{Traditional IP realization using OBC-DA for $N=4$.}\label{fig2}
		\end{figure}
		From (\ref{eq14d}), it is evident that the term ${{d}}^{ij}_{m,k}$ can take one of $2^{q-1}$ possible combinations, since $\Delta u_{{l_{mn}^{(i-j)},k}} \in \{-1, 1\}$ for $1 \leq m, n \leq q$ and $1 \leq i,j \leq p$. These combinations can be pre-computed and stored in a LUT, as shown for $q=N=4$ in Fig.~\ref{fig2}. 
		{These LUT contents correspond to the first row of the circulant MVM matrix and are accessed using the input bit-slices $(u_{n,B-1}, \ldots, u_{n,0};\ 1\leq n \leq N)$ as addresses, with bit-slices supplied in order from the least significant bit (LSB) to the most significant bit (MSB). During each access, the LUT output undergoes an SA operation, and this process continues for $B$ clock cycles to produce the output $v_1$, where $S_1$ and $S_2$ are the control signals used to load $d_{ot}$ and negate the LUT output, corresponding to the MSB and LSB of the inputs, respectively.} {Note that, as shown in (\ref{eq14dd}), and since each row of the circulant matrix contains the same set of weights, only cyclically reordered, this offset $d_{ot}$ takes the same value for every row.}
		
		\begin{figure}[t]
			\centering
			\includegraphics[width=0.96\linewidth]{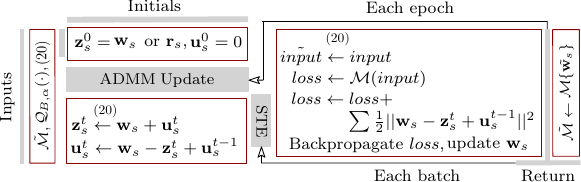}
			\caption{Illustration of the FxP training strategy for the compressed LSTM.}\label{fig:flowchart}
		\end{figure}
		\subsection{Training Strategy for Reformulated LSTM}
		\label{sec:timit_training}
		
		Quantization complicates backpropagation, as discrete values can make gradients undefined \cite{wang2017accelerating}. This is addressed using STE \cite{chang2021mix}, which treats the quantizer as the identity during backpropagation, and ADMM \cite{leng2018extremely}, which formulates quantization as a constrained optimization problem and avoids backpropagating through the quantizer by decoupling constraints via auxiliary variables and projection. ADMM is adopted to quantize the parameter vectors, as it enforces the quantization constraint more rigorously and yields improved accuracy retention for weights, while STE is used to quantize the activations. This produces the quantized LSTM model, as shown in Fig.~\ref{fig:flowchart}, achieving effective compression while preserving accuracy.
		
		During the training phase, ${\mathbf{w}}_s$, ${\mathbf{r}}_s$, and ${\mathbf{b}}_s$ are learned. The LSTM model outlined in (\ref{eq8}), (2)--(3) was initially trained using 32-bit floating-point (FlP) model $(\mathcal{M})$. This ensures that, after training, the LSTM can be efficiently executed with a lower fixed-point (FxP) model $(\mathcal{\tilde{M}})$ on the FPGA. The relevant parameters, including ${\bf{w}}_s$, ${\bf{r}}_s$, and ${\bf{b}}_s$, are quantized to a target bit-width $B$. Under FxP quantization, the values are scaled by a factor $\alpha$, and the quantized form is
		\begin{equation}
			\mathcal{Q}_{B,\alpha}(x) =  \alpha' \cdot \text{round}\left(x \cdot (2^{B-1} - 1)\right)
		\end{equation}
		where $\alpha'=\alpha/(2^{B-1} - 1)$ is the normalized $\alpha$. Consider the $B$-bit weight $\tilde{w}_{r_n}$, projected from ${\mathcal{M}}\{{w}_{r_n}\}$, as per
		\begin{equation}
			\hspace{-0.2cm}\tilde{w}_{r_n} = \alpha \cdot \text{round}\left(\max(-1, \min({w}_{r_n}/\alpha, 1))\right)
		\end{equation}
		This step combines two processes: clipping and rounding. First, ${w}_{r_n}$ is clipped to the range $[-\alpha, \alpha]$ to prevent values from exceeding the FxP limits. The clipped value is then scaled by the factor for a chosen $B$, rounded to the nearest integer, and rescaled back to the original range by $\alpha$.

		LSTMs typically have limited tolerance to numerical error, primarily due to their reliance on approximating the nonlinear activation function $\phi(\cdot) \in \{\sigma(\cdot), \tanh(\cdot)\}$ with piecewise linear functions. Despite their nonlinearity, these functions can be implemented efficiently using combinational logic gates and an LUT, e.g., $\tanh(\cdot)$, similar to \cite{zamanlooy2013efficient}; the same approach is extended to $\sigma(\cdot)$ using the identity $\sigma(x) = (1 + \tanh(x/2))/2$. The forward pass uses these piecewise linear approximations, while the backward pass computes gradients using the original nonlinear functions, as in (4).

		\begin{table}[t] 
			\caption{Training Configurations and Accuracy-Compression\\ Tradeoff: TIMIT {vs. LibriSpeech-100h}}
			\label{table_merged_full} 
			\centering 
			\resizebox{1\linewidth}{!}{% 
				\begin{threeparttable} 
					\begin{tabular}{|c|c|c|c|c|c|c|c|} 
						\hline\hline 
						\multirow{2}{*}{Case} & \multirow{2}{*}{Precision} & 
						\multicolumn{2}{c|}{TIMIT} & 
						\multicolumn{2}{c|}{{LibriSpeech-100h}} & 
						\multicolumn{2}{c|}{CR (\%)} \\ \cline{3-8} 
						& & PER (\%) & {$\Delta$PER (\%)} & 
						{WER (\%)} & {$\Delta$WER (\%)} & 
						TIMIT & {LibriSpeech} \\ \hline\hline 
						
						a) Baseline & FlP32 
						& {23.54} & {0.00} 
						& {12.84} & {0.00} 
						& -- & -- \\ \hline 
						
						b) Circulant & FlP32$^{\ast}$ 
						& {24.67} & {+1.13} 
						& {13.58} & {+0.74} 
						& -- & -- \\ \hline 
						
						c) Circulant$^{\circ}$ & FxP8$^{\ast\ast}$ 
						& {21.64} & {--1.90} 
						& {11.45} & {--1.39} 
						& {93.64} & {93.72} \\ \hline\hline 
						
						$q=1$ & FxP8 
						& {22.05} & {--1.49} 
						& {12.09} & {--0.75} 
						& {75.00} & {75.00} \\ \hline 
						
						$q=2$ & FxP8 
						& {22.05} & {--1.49} 
						& {12.08} & {--0.76} 
						& {84.36} & {84.36} \\ \hline 
						
						$q=4$ & FxP8 
						& {22.05} & {--1.49} 
						& {12.06} & {--0.78} 
						& {89.03} & {89.03} \\ \hline 
						
						$q=8$ & FxP8 
						& {22.02} & {--1.52} 
						& {12.04} & {--0.80} 
						& {91.37} & {91.40} \\ \hline 
						
						$q=16$ & FxP8 
						& {21.97} & {--1.57} 
						& {11.99} & {--0.85} 
						& {92.54} & {92.57} \\ \hline 
						
						$q=32$ & FxP8 
						& {21.91} & {--1.63} 
						& {11.93} & {--0.91} 
						& {93.13} & {93.16} \\ \hline 
						
						$q=64$ & FxP8 
						& {21.84} & {--1.70} 
						& {11.86} & {--0.98} 
						& {93.42} & {93.45} \\ \hline\hline

					\end{tabular} 
					\begin{tablenotes} 
						\item LEGEND: $^{\circ}$retrained from b); $^{\ast}$circulant weight matrices; $^{\ast\ast}$weight vectors; Block-circulant ($q$-sweep) rows are independently retrained from case b), varying circulant block size $q$, under the same FxP8/ADMM/STE procedure as case c). {CR: $1-\frac{B(3N/q+N+2)}{64(2N+1)}$ for the $q$-sweep}, and $1-\frac{B(N+5)}{64(2N+1)}$ for case c); $\Delta$PER/$\Delta$WER are relative to case a). TIMIT uses $N=256$; LibriSpeech-100h uses $N=1024$. 
					\end{tablenotes} 
				\end{threeparttable} 
			} 
		\end{table}
		
		\subsubsection{Validation on TIMIT and LibriSpeech}
		The training strategy is validated on two speech recognition benchmarks, TIMIT \cite{garofolo1993darpa}
		and LibriSpeech \cite{panayotov2015librispeech}, to confirm the accuracy-compression tradeoff generalizes
		beyond a single, small corpus. TIMIT comprises 630 speakers across eight American English dialects, each
		reading ten phonetically rich sentences, totaling roughly 500 hours of training, 50 hours of validation,
		and 5 hours of test data. Raw audio is converted to spectrograms using a Hamming window with 50\%
		overlap. The acoustic model is implemented in PyTorch, comprising convolutional layers, an LSTM layer,
		and a fully connected layer, with batch size set to the kernel size. Performance is measured by phone
		error rate (PER) \cite{li2018efficient}, with a unidirectional $N=256$ LSTM layer used throughout
		Table~\ref{table_merged_full}.
		
		{LibriSpeech is a substantially larger audiobook-speech corpus totaling 1000 hours; the standard
			100-hour training subset is used here, following the training and evaluation protocol of the
			PyTorch-Kaldi baseline \cite{ravanelli2019pytorch}. Acoustic features are 40-dimensional fMLLR features
			extracted with the Kaldi toolkit \cite{povey2011kaldi}, and decoding uses a Viterbi decoder with a
			4-gram language model. Performance is measured by word error rate (WER). LibriSpeech uses a
			$N=1024$ LSTM layer, four times larger than TIMIT, matching layer dimensions used in
			large-vocabulary LSTM accelerators \cite{li2019rnn, cao2019efficient, wang2019lstm, gao2022spartus,
				li2023fpga, kim2023v, kim2024auto}, testing whether the accuracy-compression relationship established at
			$N=256$ holds at a scale where each circulant block spans a much larger portion of the weight matrix. A
			unidirectional LSTM is used for both datasets, consistent with the low-latency, streaming target of the
			circulant/block-circulant hardware accelerator, since a bidirectional layer requires the complete utterance before
			producing any output -- a known accuracy cost \cite{graves2005bidirectional} (as also noted in \cite{gao2022spartus}) necessary for real-time
			deployment, so results here are not directly comparable to bidirectional baselines such as
			\cite{ravanelli2019pytorch}. This work's objective is to isolate the accuracy cost of the block-circulant
			constraint against a matched, consistently trained baseline, rather than to pursue state-of-the-art
			absolute accuracy.}
		
		{In both cases, an initialization similar to \cite{wang2017accelerating} preserves the variance of the
			original weights when constructing circulant matrices, stabilizing training under the structural
			constraint. FxP training combines ADMM for parameter quantization and STE for activation
			quantization, with re-training used to mitigate random-initialization effects; the block-circulant
			parameterization (9) follows standard backpropagation \cite{li2018efficient} in both cases. Both parameters and activations are quantized to FxP8 in each dataset, as lower bit-widths worsen performance, isolating the effect of dataset scale
			and block size $q$ from numerical precision. The training schedules differ, however: TIMIT
			pretrains for 100 epochs before FxP retraining, while LibriSpeech-100h pretrains for 20 epochs, with FxP
			retraining under each $q$ converging in only 10 epochs, reflecting faster convergence from a
			better-conditioned FlP32 initialization and the curriculum warm-start from the previous block size's
			converged weights.}
		
		{For TIMIT, Table~\ref{table_merged_full} reports three training cases: a) a floating-point (FlP32)
			baseline with standard square weight matrices and exact nonlinear activations, as in (\ref{eq4}); b)
			FlP32 with circulant weight matrices (5) and approximate nonlinear activations
			\cite{zamanlooy2013efficient}; and c) FxP8 parameters and activations with circulant matrices as in (6),
			retrained from b). The same cases are reported for LibriSpeech under identical precision settings. This
			retraining in c) achieves a 1.90\% PER improvement over baseline a) on TIMIT, motivating the choice of
			FxP8 for hardware execution (compression ratio of 93.64\%).}
		
		{Block-circulant retraining follows a curriculum schedule: each block size $q$ is warm-started from the
			converged weights of the previous, smaller $q$, with an increased number of ADMM iterations relative to
			case c), progressively enforcing the structural constraint rather than imposing it in one step. Since $q{=}N$ is structurally equivalent to full-circulant, both CR and PER converge to case c)'s values as $q\to N$, consistent with $q{=}64$'s CR (93.42\%) and PER (21.84\%)
			approaching case c)'s CR (93.64\%) and PER (21.64\%) at $q{=}N{=}256$.}
		
		{As Table~\ref{table_merged_full} shows, this convergence pattern holds for both datasets despite
			LibriSpeech-100h providing roughly $5\times$ less training data than TIMIT (100h versus 500h) and
			using a fourfold larger layer ($N=1024$ versus $N=256$). That comparable stability persists despite less
			data and a larger layer suggests the block-circulant structure's own regularizing effect, rather than
			dataset scale, is the dominant factor in the observed accuracy-compression tradeoff. The LibriSpeech
			evaluation, using a pipeline directly derived from \cite{ravanelli2019pytorch}, thus exhibits the same
			qualitative accuracy-compression relationship as TIMIT, supporting the conclusion that this accuracy cost
			is not tied to the particular baseline architecture used in the TIMIT experiments.}
		
		{Using FxP8 at $N=1024$ is a controlled but nontrivial choice: larger hidden-state dimensions typically
			accumulate wider-dynamic-range activations per gate (particularly the cell state $c^t$) than the
			256-unit TIMIT configuration, so FxP8 may quantize LibriSpeech activations more coarsely relative to
			their true range. LibriSpeech WER indeed degrades roughly twice as fast, in relative terms, as
			TIMIT PER across the $q$-sweep, consistent with this
			activation-precision effect compounding the block-circulant structural cost at larger scale, rather than
			indicating a genuine weakness of the block-circulant structure itself. The ADMM/STE retraining procedure, which adapts the network to the reduced-precision representation rather than quantizing post hoc, partially mitigates this effect.}

		\section{Hardware Optimization and  Accelerator}

		\begin{figure}[t]
			\centering  \includegraphics[width=0.88\linewidth]{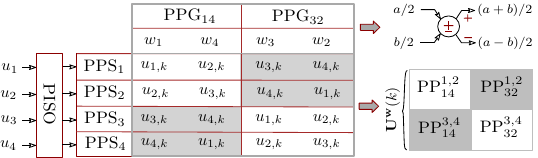}
			\caption{Proposed initial scheme for circulant MVM with $N=4$.}\label{fig5}
		\end{figure}
		\begin{figure}[t]
			\centering  \includegraphics[width=0.84\linewidth]{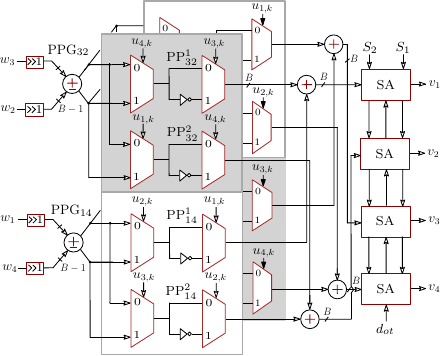}
			\caption{Proposed initial circulant MVM architecture with $N=4$.}\label{fig6}
		\end{figure}
		
		\subsection{Initial Scheme and Architecture}
		The proposed scheme leverages OBC to generate and select PPs in pairs. For instance, a pair of weights \(a\) and \(b\) forms the PPs \((\pm a \pm b)/2\), which possess the anti-symmetry property, i.e., \(-(a+b)/2 \leftrightarrow (a+b)/2\) and \(-(a-b)/2 \leftrightarrow (a-b)/2\) are anti-symmetric pairs. This means that only half of the combinations need to be generated, while the other half can be obtained through negation. Fig.~\ref{fig5} illustrates the proposed scheme with two PPGs, PPG$_{14}$ and PPG$_{32}$, for $N=4$, corresponding to weight pairs $\{w_1,w_4\}$ and $\{w_3,w_2\}$. The subscripts denote their associated weight pairs. The architecture uses the circulant input matrix $\mathbf{U}=\{u_{mn}\}_{m,n=1}^{N}$, where $mn={(m+n-2)}\bmod(N)+1$. Each $B$-bit input $u_{mn}$ is bit-sliced into $u_{mn,k}$ ($0 \leq k \leq B-1$) via a PISO converter~\eqref{eq14}, serving as select signals in ${\mathbf{U}}^{\mathbf{w}}(k)$ for PPG$_{14}$ and PPG$_{32}$. Here, the superscript ${\mathbf{w}}$ indicates PPs formed from the weight vector and input slices, and $k$ denotes the index of input bit-slice. The PPGs/PPSs generate a $2\times 2$ chessboard pattern in ${\mathbf{U}}^{\mathbf{w}}(k)$, dividing PPSs into two vertical groups. White regions contain PP$^m_{14}|_{m=1,2}$ and PP$^m_{32}|_{m=3,4}$, while gray regions contain PP$^m_{32}|_{m=1,2}$ and PP$^m_{14}|_{m=3,4}$. Although elements in each division appear identical, they cannot be shared, as each selects a different PP depending on its position.

		\begin{figure}[t]
			\centering  \includegraphics[width=0.74\linewidth]{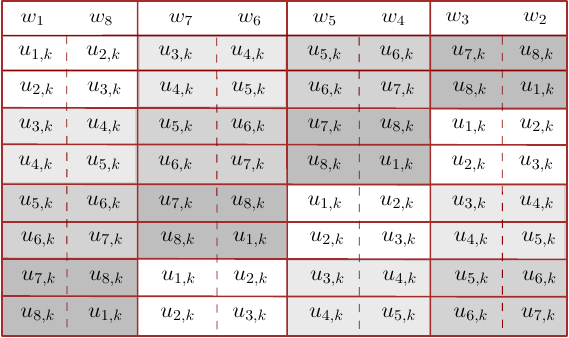}
			\caption{Proposed initial scheme for circulant MVM with $N=8$.}\label{fig7}
		\end{figure}
		
		The initial hardware LUT architecture for $N=4$, employing PPGs/PPSs to compute the circulant MVM, is shown in Fig.~\ref{fig6}. The first-row output $v_1$ is computed using PP$_{14}^1$ and PP$_{32}^1$ from ${\mathbf{U}}^{\mathbf{w}}(k)$. For PP$_{32}^1$, PPG$_{32}$ generates $(w_3 \pm w_2)/2$, selected through two 2-to-1 multiplexers: the first controlled by $u_{4,k}$ and the second by $u_{3,k}$, which applies a TC operation when required. Similarly, PPG$_{14}$ produces $(w_1 \pm w_4)/2$ in PP$_{14}^1$ with its own two-stage selection. Outputs from PP$_{14}^1$ and PP$_{32}^1$ are summed and processed by the SA unit to yield $v_1$. Multiplexers of the same color correspond to chessboard divisions. The same principle extends to compute other row outputs $v_m$ using the PP$^m_{14}$ and PP$^m_{32}$.

		The design scales to larger $N$ by extending the chessboard pattern. For $N=8$, pairing weights in groups of $N/2$ yields $2\times2$ divisions of size $4\times4$, which increases PPG/PPS complexity. Each PPG must generate 16 combinations of $(\pm a \pm b \pm c \pm d)/2$, of which only half are required, using 12 adders/subtractors and two PPS units, each with eight 8-to-1 multiplexers. A more efficient alternative is pairing by $N/4$, which preserves the $N=4$ structure, as shown in Fig.~\ref{fig7}. This requires only four PPGs (one adder/subtractor each) and four PPSs (each with four two-level 2-to-1 multiplexers). In general, an $N \times N$ circulant MVM scales into $2 \times 2$ chessboard divisions, requiring $N/2$ PPGs {(each containing an adder/subtractor)} and {$(N/2)^2$ PPSs (each containing 4 2-to-1 multiplexers).}

		A natural question arises: why use circulant input matrices for the MVM instead of circulant weight matrices? In LSTMs, this choice is advantageous for PPG/PPS-based designs because PP generation is independent of selection, enabling bit-level optimizations. Because an LSTM requires $8N$ weight parameters but only $2N$ inputs, input-circulant MVMs reduce hardware cost significantly. Weight/recurrent-weight circulant MVMs need $8(N/2)$ PPGs and $8(N/2)^2$ PPSs, whereas input-circulant MVMs require only $2(N/2)$ PPGs (for inputs and recurrent inputs), cutting PPG count by $4\times$ and reducing PPS demand. Since a PPG with a conditional adder is costlier than a PPS multiplexer, this motivated reformulating circulant MVM into matrix (input)--vector (weight) MVM. Moreover, the input ($\mathbf{X}^t\mathbf{w}_s$) and recurrent ($\mathbf{Y}^{t-1}\mathbf{r}_s$) circulant MVMs share the same computational structure and are time-multiplexed, reusing the same PPG/PPS set rather than duplicating hardware for each.

		%\section{Optimized Architecture}
		\subsection{{PPG/PPS Optimization through PPs Column Swapping}}
\begin{figure}[t]
			\centering  \includegraphics[width=0.92\linewidth]{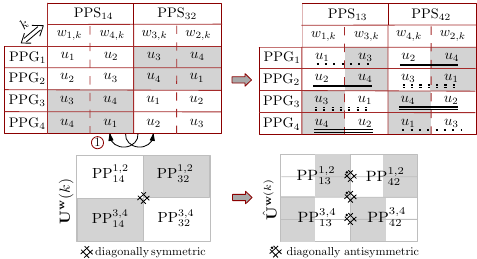}
			\caption{PPG/PPS optimization for $N=4$ circulant MVM {via one-time input-PP column swapping}.}\label{fig8}
		\end{figure}

		As noted earlier, PPG and PPS complexities scale linearly and quadratically with \(N/2\), respectively, with PPSs becoming the main bottleneck for compressed LSTMs at larger \(N\). Reducing PPS complexity is therefore critical. One approach is to apply bit-slice operations via PISO to the weights rather than the inputs, modifying the PPs matrix from ${\mathbf{U}}^{\mathbf{w}}(k)$ to ${\mathbf{U}}^{\mathbf{w}(k)}$ with elements PP$_{r_{n}r_{n+1}}^m$, where $r_n=(1-n)\bmod(N)+1$. This enables interchange of PPGs and PPSs, as shown for $N=4$ in Fig.~\ref{fig8}. In this scheme, PPs are generated by PPGs, requiring two sets of four PPGs, with bit-sliced weight pairs selected in the PPSs. Unlike the initial scheme, which shares two PPGs across all PPSs, this approach applies the chessboard division to PP generation rather than PP selection. However, it doubles the PPG count to four due to diagonal symmetry, motivating further structural reformulation to reduce hardware cost.

\begin{figure}[t]
			\centering  \includegraphics[width=0.85\linewidth]{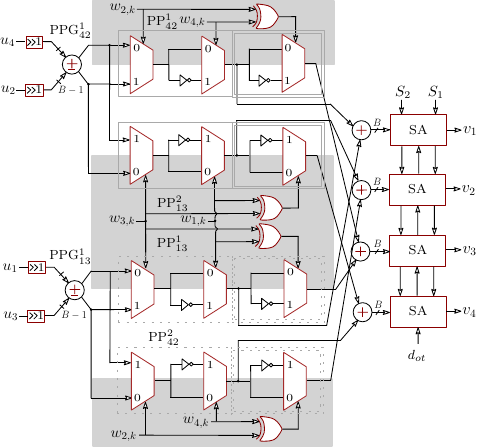}
			\caption{Column-swapped circulant MVM architecture for $N=4$.}\label{fig9}
		\end{figure}

		The four PPGs in this approach result from element repetition within each chessboard division. For instance, PP$_{14}^{1}$ and PP$_{14}^{2}$ duplicate $u_2$ along the anti-diagonal, limiting PPG/PPS optimization. {To address this, the chessboard pattern of the PP-matrix ${{\mathbf{U}}^{{\mathbf{w}}(k)}}$ is altered by a one-time column swap of the input PPs for the unpaired bit-slices $w_{4,k}$ (Col 2) and $w_{3,k}$ (Col 3)}, as shown in step \circled{1} of Fig.~\ref{fig8}, yielding a new PPs matrix ${\hat{\mathbf{U}}^{{\mathbf{w}}(k)}}$. This swapping of PPs is possible since summation is order-independent and thus does not affect the output, as per (\ref{eq14dd}). Re-pairing the swapped bit-sliced weights into $\{w_{1,k},w_{3,k}\}$ and $\{w_{4,k},w_{2,k}\}$ forms new PPG/PPS sets. Consequently, only $(\pm u_1 \pm u_3)/2$ and $(\pm u_2 \pm u_4)/2$ need to be generated for Row 1, while the corresponding combinations for Row 3, $(\pm u_3 \pm u_1)/2$ and $(\pm u_4 \pm u_2)/2$, are redundant due to diagonal symmetry or anti-symmetry. For example, bit combination $00$ produces the same value, $(u_1+u_3)/2$, for both $(\pm u_1 \pm u_3)/2$ and $(\pm u_3 \pm u_1)/2$, whereas $01$ yields their anti-symmetric counterpart, $\pm(u_1-u_3)/2$. Similar behavior occurs for combinations $10$ and $11$. Consequently, a bit-level XOR of the $01$ and $10$ cases can serve as the select signal for a 2-to-1 multiplexer to obtain the TC versions of the PPs. This reduces the PPG count to two (PPG$_{14}^{1}$ and PPG$_{32}^{1}$), as in the initial scheme. Distinct PPG inputs are marked with a single underline in Fig.~\ref{fig8}, while diagonally anti-symmetric elements use a double underline. The new architecture uses PPG$_{42}^1$ to generate $(u_2\pm u_4)/2$, selected by $w_{2,k}$, and PPG$_{13}^1$ to generate $(u_1\pm u_3)/2$, selected by $w_{3,k}$. The column-swapped architecture of circulant MVM for $N=4$, shown in Fig.~\ref{fig9}, halves the PPS count, with minor added cost from a few 2-to-1 multiplexers handling anti-diagonal symmetric PPs. {It is observed that each pair of PPs and their corresponding TC versions requires a dedicated adder to compute the PPs sum for a desired row prior to the SA unit.}

		\begin{figure}[t]
			\centering  \includegraphics[width=0.98\linewidth]{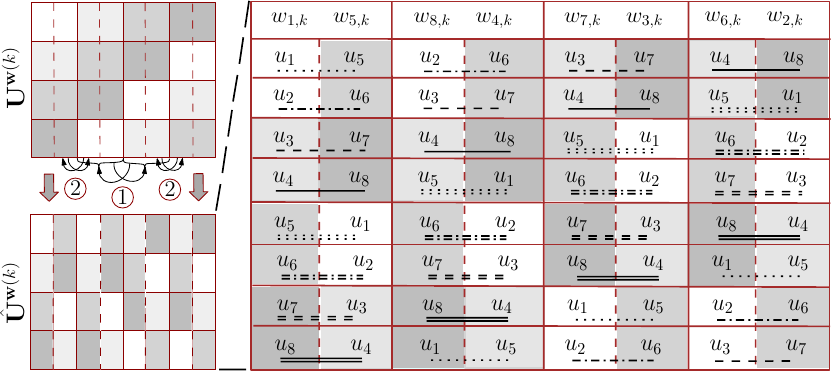}
			\caption{PPG/PPS optimization for $N=4$ circulant MVM {via two-times input-PP column swapping.}}\label{fig10}
		\end{figure}

		For $N=8$, only four PPGs are needed to generate $(\pm u_1 \pm u_5)/2$, $(\pm u_2 \pm u_6)/2$, $(\pm u_3 \pm u_7)/2$, and $(\pm u_4 \pm u_8)/2$, while the remaining components (double underlined) are redundant. This yields an optimized circulant MVM architecture for $N=8$, similar to Fig.~\ref{fig9}. Unlike $N=4$, where the column swap occurs only once at $N/2$, for $N=8$ the process unfolds in two steps, requiring two swaps. In step~\circled{1}, columns are swapped at $N/2$ between adjacent left/right groups with swap width (SW) $N/4$. In step~\circled{2}, two smaller swaps occur at $N/4$ ($=N/2-N/4$) and $3N/4$ ($=N/2+N/4$) with SW $N/8$, as shown in Fig.~\ref{fig10}. Extending this approach reveals a regular pattern: column-swapping permutations depend on step level, SW, number of swaps (NOS), and swap positions (POS). The elements of $\hat{\mathbf{U}}^{\mathbf{w}}(k)$, i.e., PP$^{r_{\pi(n)}r_{\pi(n)+1}}_m$, with permutations $r_{\pi(n)}$, are obtained using Algorithm~1. Any $N \times N$ circulant MVM can thus be implemented with the optimized LUT-based architecture using $2 \times 2$ chessboard divisions, requiring $N/2$ PPGs and $2(N/4)^2$ PPSs. While the PPG count remains unchanged, the optimized scheme reduces the total PPS requirement for $\mathbf{w}_s/\mathbf{r}_s$ to $(N/2)^2 + 3 \cdot 2(N/4)^2$ {(with circulant MVM each now containing 6 2-to-1 multiplexers)}, yielding a 37.5\% reduction compared to the initial design.

		\begin{figure}[t]
			\centering
			\scalebox{0.9}{
				\begin{minipage}{\linewidth}
					\begin{algorithm}[H]
						\caption{PPs Column-Swapping for Circulant MVMs}
						\begin{algorithmic}[1]
							\STATE \textbf{Input:} PPs matrix ${\mathbf{U}^{\mathbf{w}(k)}}$, $r_n=(1-n)\bmod (N)+1$
							\STATE \textbf{Output:} Permuted PPs ${\hat{\mathbf{U}}^{\mathbf{w}(k)}}$, $r_{\pi(n)}$
							\STATE Initialize ${\hat{\mathbf{U}}^{\mathbf{w}(k)}} \gets {\mathbf{U}^{\mathbf{w}(k)}}$, $r_{\pi(n)} \gets r_n$
							\FOR{$s = 1$ to $\log_2 N - 1$}
							\STATE $\text{SW} \gets N/2^{s+1}$, $\text{NOS} \gets 2^{s-1}$
							\FOR{$i = 1$ to $\text{NOS}$}
							\STATE $\text{POS} \gets ((2i-1)N)/2^s$
							\STATE $\text{left} \gets \text{POS}-\text{SW}+1:\text{POS}$
							\STATE $\text{right} \gets \text{POS}+1:\text{POS}+\text{SW}$
							\STATE Swap columns: ${\hat{\mathbf{U}}^{\mathbf{w}(k)}}[:,\text{left}] \leftrightarrow {\hat{\mathbf{U}}^{\mathbf{w}(k)}}[:,\text{right}]$
							\STATE Swap indices: $r_{\pi(n)}[\text{left}] \leftrightarrow r_{\pi(n)}[\text{right}]$
							\ENDFOR
							\ENDFOR
							\STATE \textbf{return} $({\hat{\mathbf{U}}^{\mathbf{w}(k)}}, r_{\pi(n)})$
						\end{algorithmic}
					\end{algorithm}
				\end{minipage}
			}
		\end{figure}

		\subsection{{Optimized-Complexity Design of Circulant MVM}}
\begin{figure}[t]
			\centering  \includegraphics[width=1\linewidth]{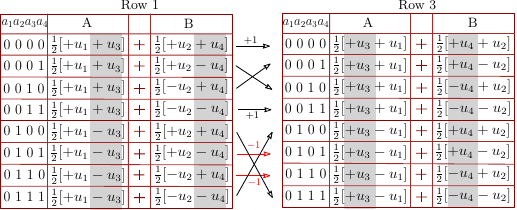}
			\caption{{Mapping of Row 1 PPs in pair to Row 3 PPs in pair for $N=4$.}}\label{fig12n}
		\end{figure}

		{As shown in Fig.~\ref{fig12n}, analysis of input PPs in pairs for a circulant MVM at $N=4$ reveals redundancy between Rows~1 and~3. Each row's PPs sum is formed from two PPs in a pair, denoted $(\text{A,B})$ below, yielding 16 combinations, of which only 8 are antisymmetric. Comparing Row~1 and Row~3 entry-by-entry, three cases emerge: some addresses (e.g., 0000, 0011) give identical $(\text{A,B})$ pairs; others (e.g., 0101, 0110) give exact negations; and the rest (e.g., 0001, 0010, 0100, 0111) match neither, since the two terms in $(\text{A,B})$ diverge independently. Only the first two cases permit reuse of Row~1's already-computed PPs in pair through negation alone; the remaining cases require an explicit correction. For simplicity, only the half-combination pair A$_{\text{h}}=(u_1 \pm u_3)/2$ and B$_{\text{h}}=(u_2 \pm u_4)/2$ is considered here, where subscript `h' denotes this half-set. Row~1's per-cycle PPs sum, Su$_1$, together with its antisymmetric version, is accumulated over $B$ bit-slice cycles by its SA unit,}
		\begin{equation}
			{\text{SA}_1[k] = \text{Su}_1[k] \pm \tfrac{1}{2}\,\text{SA}_1[k-1], \qquad \text{SA}_1[-1]=d_{ot},} \label{eq:acc1n}
		\end{equation}
		{where $k$ denotes the clock cycle index of the SA unit and `$-$' in $\pm$ corresponds to the MSB of the weights. This operation is identical to the SA unit shown in Fig.~\ref{fig2}. Row~3's PP sum ($\text{Su}_3$) requires an explicit adder for $A_{\text{h}}=(u_3 \pm u_1)/2$ and $B_{\text{h}}=(u_4 \pm u_2)/2$, which are Row~1's terms with signs exchanged for $01$ and $10$ cases, as shown in Fig.~\ref{fig9}. Therefore, Row~3's PPs sum can be expressed relative to Row~1's as}
		\begin{equation}
			{\text{Su}_3[k] = \text{Su}_1[k] - \Delta_3[k], } \label{eq:row3_correction}
		\end{equation}
		{where $\Delta_3[k]$ is the correction factor that obtains $\text{Su}_3[k]$ from $\text{Su}_1[k]$. It is $0$ when the weight-bit pair governing Row~1 and Row~3 agree at clock cycle $k$. In this case, Row~3's PPs sum is tapped directly (or negated) before entering Row~1's SA feedback, and is combined with $\Delta_3[k]=0$ to form $\text{Su}_3[k]$. Otherwise, $\Delta_3[k] \in \big\{(u_1-u_3),\ (u_2-u_4)\big\}$ is nothing but a left-shift on the generated input PPs, obtained directly by comparing Row~3's pair of PPs with Row~1's pair of PPs.}
		
		{Because Row~1's SA accumulates recursively via \eqref{eq:acc1n} as per \eqref{eq14dd}, Row~3's accumulated output must satisfy the same recursion with $\text{Su}_3$ in place of $\text{Su}_1$. The running correction to Row~1's accumulation, $\text{CB}_3 \triangleq \text{SA}_1-\text{SA}_3$, is realised by a lightweight correction unit termed a Correction Buffer (CB), and therefore obeys the same recursive form,}
		\begin{equation}
			{\text{CB}_3[k] = \Delta_3[k] + \tfrac{1}{2}\,\text{CB}_3[k-1],} \label{eq:corr_recursion}
		\end{equation}
		{so a feedback-free register holding only the current correction cannot reproduce this recursive decay, and is therefore insufficient once mismatches occur more than once. The CB retains the shift-feedback structure of an SA unit but requires no dedicated adder to produce $\text{Su}_3$, since its input is drawn entirely from the correction factor $\Delta_3[k]$.}
		
		{The equal (EQ) and correction-select (CS) control signals for the correction factors are derived from the weight bits at negligible cost:}
		\begin{align}
			{\text{EQ}} & {=1 \oplus w_{1,k} \oplus w_{3,k} \oplus w_{4,k} \oplus w_{2,k},} & {\text{CS}} & {=w_{1,k} \oplus w_{3,k}.}  \label{eq:cf} \nonumber
		\end{align}
		{where `$\oplus$' exclusive OR (XOR) operator. $\text{EQ}=1$ indicates agreement, so $\Delta_3[k]=0$ and Row~1's tapped operand passes through unmodified. When $\text{EQ}=0$, the two pair-terms are forced into disagreement with one another, so a single XOR suffices to identify which term needs correcting: since $\text{EQ}=0$ implies $w_{1,k}\oplus w_{3,k}$ and $w_{4,k}\oplus w_{2,k}$ are complementary, the mismatch is confined to exactly one of the two pair-terms, and testing $w_{1,k}\oplus w_{3,k}$ alone identifies which one it is. $\text{CS}$ selects $\Delta_3[k]=(u_1-u_3)$ if $\text{CS}=1$, or $\Delta_3[k]=(u_2-u_4)$ if $\text{CS}=0$.}

\begin{figure}[t]
			\centering  \includegraphics[width=0.92\linewidth]{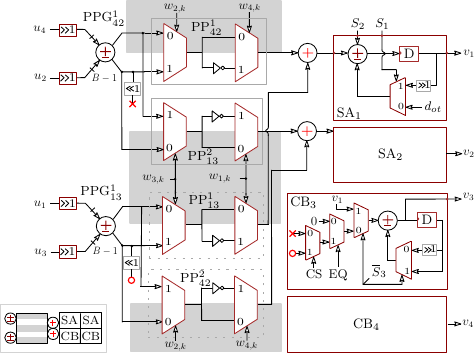}
			\caption{{Optimized-complexity circulant MVM architecture for $N=4$ with its pictograph, where $\times: u_2-u_4$ and $\circ: u_1-u_3$.}}\label{fig9n}
		\end{figure}

		{At the $B$-th (final) clock edge, $\text{SA}_1[B{-}1]$ and $\text{CB}_3[B{-}1]$ -- computed during cycle $k=B{-}1$ -- are available on their respective registers. (This contrasts with the preload $\text{SA}_1[-1]=d_{ot}$ in \eqref{eq:acc1n}, set before accumulation starts, i.e., before $k=0$.) Unlike $\text{SA}_1$, $\text{CB}_3$ carries no offset, since the offset cancels in the difference $\text{CB}_3 = \text{SA}_1 - \text{SA}_3$; this leaves its adder free for reuse. Once cycle $k=B{-}1$ completes, the control signal $\overline{S}_3$ switches, at the falling clock edge, the CB unit's adder input away from the correction term $\Delta_3$ and onto the tapped operand $\text{SA}_1[B{-}1]$, so the same adder that accumulates $\text{CB}_3$ during correction now forms $\text{SA}_1[B{-}1]-\text{CB}_3[B{-}1]$ ahead of the register's (rising-edge) capture:}
		\begin{equation}
			{v_3 = \text{SA}_1[B{-}1] - \text{CB}_3[B{-}1]} \label{eq:final_combine}
		\end{equation}
		{with $\overline{S}_3$ deasserted again at the start of the next sample, returning the adder to its correction-accumulation role -- no preload is required here, in contrast to $\text{SA}_1$'s $d_{ot}$ load.
			Fig.~\ref{fig9n} shows the resulting datapath. Two 2-to-1 multiplexers, controlled by $\text{EQ}$ and $\text{CS}$, select $\Delta_3[k]$ for $k=0,\dots,B{-}1$.
			The identical approach, applied with the complementary term assignment, produces Row~4 from Row~2 via its own $\text{CB}_4$. Row~4's correction factor is likewise $0$ when the weight-bit pair governing Row~2 and Row~4 agree, and otherwise $\Delta_4[k] \in \big\{(u_4-u_2),\ (u_3-u_1)\big\}$ i.e., the negatives of Row~3's nonzero correction values; Row~4's CB therefore reuses the same magnitudes already generated from the input PPs as Row~3, requiring only an inverted sign selection. This optimized version of circulant MVM, shown in Fig.~\ref{fig9n}, has two fewer adders for the PPs sum, and each PPS contains only 4 2-to-1 multiplexers, compared to Fig.~\ref{fig9}.}
		
		\subsection{{Optimized-Complexity Design for Higher Model Size}}
		
		{For $N=8$, Row~1's sum is formed from four PP pairs, A$_{\text{h}}=(u_1\pm u_5)/2$,
			B$_{\text{h}}=(u_2\pm u_6)/2$, C$_{\text{h}}=(u_3\pm u_7)/2$, D$_{\text{h}}=(u_4\pm u_8)/2$, shared across
			all rows. Fig.~\ref{fig12n} shows Row~5's terms, $(\pm u_5\pm u_1)/2,\ (\pm u_6\pm u_2)/2,\ (\pm u_7\pm
			u_3)/2,\ (\pm u_8\pm u_4)/2$, are Row~1's four terms with each pair's roles and signs independently
			exchanged. Up to two of $\{A_{\text{h}},B_{\text{h}},C_{\text{h}},D_{\text{h}}\}$ can therefore diverge
			from Row~1 to Row~5 on the same cycle, so a single mux-selected correction no longer suffices.}
		
		{Grouping the terms as $P_{\text{h}}=A_{\text{h}}+B_{\text{h}}$, $Q_{\text{h}}=C_{\text{h}}+D_{\text{h}}$
			gives $\text{Su}_1[k]=P[k]+Q[k]$, accumulated by Row~1's SA unit as in \eqref{eq:acc1n}, and isolates two
			independent sub-pairs, each governed by its own equal/select signals in exactly the form of $N=4$. Row~5's
			correction is therefore $\Delta_5=\Delta_5^{(P)}+\Delta_5^{(Q)}$, each part $0$ on agreement and otherwise
			selected by its own mux from $\Delta_5^{(P)}\in \big\{(u_1-u_5),\ (u_2-u_6)\big\}$ and $\Delta_5^{(Q)}\in
			\big\{(u_1-u_5),\ (u_2-u_6)\big\}$.}
		%\begin{align}
		%{\Delta_5^{(P)}} &{\in \big\{(u_1-u_5),\ (u_2-u_6)\big\},} %\label{eq:deltaP}\\
		%{\Delta_5^{(Q)}} &{\in \big\{(u_3-u_7),\ (u_4-u_8)\big\},} \label{eq:deltaQ}
		%\end{align}
		{Because $(A_{\text{h}},B_{\text{h}})$ and $(C_{\text{h}},D_{\text{h}})$ are governed by disjoint
			weight bits, their agreement states are independent, so each requires its own equal/select pair:}
		\begin{align}
			{\text{EQ}_P} & {= 1 \oplus w_{1,k} \oplus w_{5,k} \oplus w_{2,k} \oplus w_{6,k},} & {\text{CS}_P} &{= w_{1,k} \oplus w_{5,k},} \nonumber\\
			{\text{EQ}_Q} & {= 1 \oplus w_{3,k} \oplus w_{7,k} \oplus w_{4,k} \oplus w_{8,k},} & {\text{CS}_Q} &{= w_{3,k} \oplus w_{7,k},} \nonumber
		\end{align}
		{$\text{EQ}_P$ and $\text{EQ}_Q$ need not agree, so $\Delta_5^{(P)}$ and $\Delta_5^{(Q)}$ can both be
			nonzero on the same cycle, and combining them requires a real addition rather than a further mux.}
		
		{A single CB is retained by summing the two mux outputs ahead of its feedback add:}
		\begin{equation}
			{\text{CB}[k] = \big(\Delta_5^{(P)}[k]+\Delta_5^{(Q)}[k]\big) + \tfrac12\,\text{CB}[k-1].} \label{eq:cb_merge_n8}
		\end{equation}
		{This merge adder is the only per-cycle arithmetic cost beyond the mux-only selection stage; it can be
			realized as one three-operand adder or two cascaded two-operand adders in the same stage.} {Merging into a single CB saves one narrow register relative to two separate CBs, and the final
			combine remains a plain two-operand subtract:}
		\begin{equation}
			{v_5 = \text{SA}_1[B{-}1] - \text{CB}[B{-}1],} \label{eq:final_combine_n8}
		\end{equation}
		{identical in form to \eqref{eq:final_combine}. $\text{CB}$ carries no per-sample offset -- it cancels
			in \eqref{eq:final_combine_n8} -- so its adder is free at the cycle where $\text{SA}_1,\text{SA}_2,\dots$
			load $d_{ot}$, provided the controller enforces $\text{EQ}_P[0]=\text{EQ}_Q[0]=1$ at the start of every
			sample; that same adder then realizes \eqref{eq:final_combine_n8} in subtract mode, with no dedicated
			final-combine adder or holding register. This trades a wider, three-operand correction-path add every
			cycle for one fewer register and a simpler final combine, compared with retaining two independent narrow,
			mux-only CBs.}
		
		{The same construction extends to the remaining row pairs, Row~2/Row~6, Row~3/Row~7, Row~4/Row~8, with
			$A_{\text{h}},B_{\text{h}},C_{\text{h}},D_{\text{h}}$ relabeled by the circulant shift. Each row-pair's
			$\Delta^{(P)},\Delta^{(Q)}$ draws a two-element subset of the same four elementary magnitude differences,
			$(u_1{-}u_5),(u_2{-}u_6),(u_3{-}u_7),(u_4{-}u_8)$, with sign selected independently per pair. Row~6's
			correction relative to Row~2, for instance, reuses $(u_2{-}u_6)$ and $(u_4{-}u_8)$ directly from Row~5's
			magnitude set and reuses $(u_1{-}u_5)$ with inverted sign. Consequently, no additional subtractor is
			required for any row-pair beyond these four elementary differences; each subsequent pair's EQ/CS logic
			only selects and sign-adjusts among the same four shared magnitudes.}
		
		{In general, for row pairs $\big(i,\ ((i-1+N/2)\bmod N)+1\big)$, the first row uses a full SA unit and
			the second uses a CB. For $N=8$, this yields four full SA units (Rows~1--4) and four CBs (Rows~5--8), each
			CB reusing the PPs sum of its paired row rather than requiring a dedicated adder. More generally, for
			$N=2^m$, each row pair decomposes into $N/4$ independent sub-corrections under repeated column swapping,
			consolidated into one CB per pair via an $(N/4)$-operand merge adder.}

		\begin{figure}[t]
			\centering  	\includegraphics[width=0.90\linewidth]{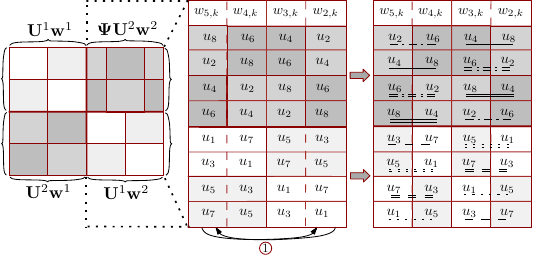}
			\caption{{PPG/PPS optimization of block-circulant MVM with $N=8$ and $p=2$ via one-time input-PP column swapping performed at the edges.}}\label{fig11}
		\end{figure}

		\subsection{Optimized-Complexity Design for Block-Circulant MVM}
		
		{Although previous optimizations apply to circulant MVMs, extending them to block-circulant MVMs requires care, as sub-input matrices ${\bf U}^{p}$ rotated by ${\bf \Psi}$ (as in (9)) disrupt the chessboard division pattern. Consider $N=8$, $p=2$, forming four sub-MVMs: ${\bf U}^{1}{\bf w}^{1}$, ${\bf U}^{2}{\bf w}^{1}$, ${\bf \Psi}{\bf U}^{2}{\bf w}^{2}$, and ${\bf U}^{1}{\bf w}^{2}$. While ${\bf U}^{1}{\bf w}^{1}$, ${\bf U}^{2}{\bf w}^{1}$, and ${\bf U}^{1}{\bf w}^{2}$ preserve the circulant chessboard pattern, ${\bf \Psi}{\bf U}^{2}{\bf w}^{2}$ is altered by the rotation, as shown in Fig.~\ref{fig11}. This prevents direct application of the column-swapping scheme. Instead of initiating column swaps at the midpoint of the PP-matrix, column swaps for the affected sub-MVMs start at the edges (see Step \circled{1} in Fig.~\ref{fig11}). For $q=4$, column swapping occurs at two positions, $q/4$ and $q$, equivalently $(q/2 - q/4)$ and $(3q/4 + q/4)$. In general, input pairs generating identical PPs can still be identified, and the scheme extends to higher $q$ by adjusting SW, NOS, and POS, following Algorithm~1.
			
			From (10), $p$ parallel sub-circulant MVMs can compute an $N \times N$ circulant MVM in $p$ clock cycles via time-multiplexing. Combining this with (11) and (18), the block-circulant MVM architecture for $q=4$ with $N=4p$ extends naturally from the optimized circulant MVM architecture for $N=4$ shown in Fig.~\ref{fig9n}, and generalizes to arbitrary $q \times q$ configurations. The summation order in (11), together with the OBC-formulated sub-MVM in (18), reduces the number of SA/CB units from $qp$ to $q$: $q/2$ full SA units (Rows~1 through $q/2$) and $q/2$ CB units (their paired rows), each a physical accumulator reused across the $p$ time-multiplexed sub-blocks, yielding substantial hardware savings.}
		
		\begin{figure}[t]
			\centering  \includegraphics[width=0.92\linewidth]{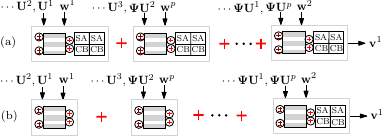}
			\caption{{Block-circulant MVM architecture for $N=4p$: (a) Initial, (b) Optimized.}}\label{fig13}
		\end{figure}
		
		{The optimized PPGs and PPSs, together with a lightweight correction scheme that reuses the paired row's PP sum to eliminate a dedicated adder for redundant-row generation, further benefit an $N\times N$ LSTM layer with block-circulant MVMs, where $N=p\times q$, particularly for $q>4$. Specifically, $q$ is scaled by a factor $\alpha$, such that $q=4\alpha$, with $p=N/(4\alpha)$; this allows the design to scale to higher $N$. For $\alpha\geq1$, column swapping of PPs together with the correction scheme requires $3\cdot2(q/4)^2=6\alpha^2$ terms, each containing 4 2-to-1 multiplexers ($24\alpha^2$ total), across the three compressed gates. Meanwhile, the number of PPGs remains $pq/2=2p\alpha$, corresponding to $4p\alpha$ adders/subtractors for generating PPs across all rows, shared across the compressed gates. Each compressed gate has its own dedicated SA/CB units, time-multiplexed across the $p$ sub-MVMs. Each gate has $2\alpha$ CB/SA units, each containing 1 adder, 1 register, and 1 2-to-1 multiplexer, giving $6\alpha$ registers and $6\alpha$ multiplexers across all three gates; the $2\alpha$ CB units additionally require $2p(\alpha-1)$ correction adders per gate, and the rows feeding the SA units contribute a further $2\alpha(p-1)$ adders per gate. Summing the correction, row-feeding, and loading adders across the three gates gives $6p\alpha(\alpha-1)+6\alpha(p-1)+6\alpha=6p\alpha^2$ adders in total.}
		
		{The block gate uses $(N/2)^2$ PPS units, each with 4 2-to-1 multiplexers, and $N$ SA units, each with 1 adder, 1 register, and 1 multiplexer. Time-multiplexed over $p$ clock cycles, its footprint reduces to one row-group of size $N/p=q$: $q$ adders, $q$ registers, and $q+pq^2$ multiplexers. Each CB unit also uses $\alpha$ CS-gated multiplexers for correction selection, contributing $2p\alpha^2$ per gate ($6p\alpha^2$ across three gates), plus two fixed $\overline{S}_3$-gated and one fixed EQ-gated multiplexer per unit ($18\alpha$ total), and $6\alpha$ SA-selection multiplexers across the three gates. For the bit-level CS logic, each sub-pair requires three XOR gates and one XNOR gate; with $2p\alpha$ sub-pairs, normalized to a word length of $B$ bits, this gives $18p\alpha/B$ XOR and $6p\alpha/B$ XNOR gates, or $24p\alpha/B$ in total. The block gate requires no CB logic, so its XOR/XNOR and correction-related terms are zero. Thus, the overall hardware complexity is $4p\alpha+6p\alpha^2+q$ adders, $6p\alpha^2+24\alpha^2+24\alpha+q+pq^2$ 2-to-1 multiplexers, $24p\alpha/B$ XOR/XNOR gates, and $12\alpha+q$ registers.}

		\begin{figure}[t]
			\centering
			\includegraphics[width=0.92\linewidth]{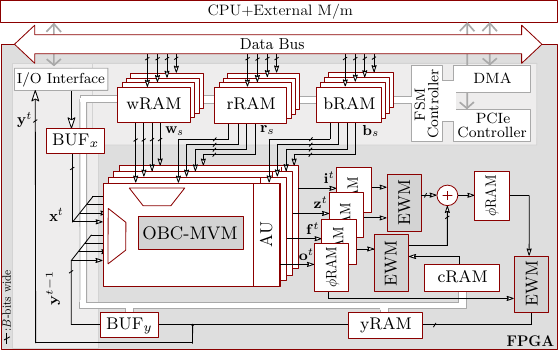}
			\caption{System-level diagram of the COBALT on FPGA.}\label{fig:system}
		\end{figure}
		
		\subsection{Proposed Hardware Accelerator: COBALT}
		The COBALT workflow consists of model development, architectural design and optimization, and hardware
		deployment. The compressed LSTM model is constructed based on the OBC scheme, followed by architecture
		optimization for efficient FPGA implementation, as discussed in the preceding subsections. This
		subsection focuses on the hardware deployment stage, which uses an AXI-based interface on the Xilinx
		ZCU104 platform: an AXI DMA engine transfers inputs, weights, and activations between external memory
		and on-chip buffers, while AXI-Lite registers handle control and synchronization with the processing
		system. All on-chip buffers use distributed RAM (built from slice LUTs) rather than dedicated BRAM,
		which reduces resource contention and keeps buffer access latency deterministic for the fine-grained,
		register-free datapath described earlier.
		\begin{figure*}[t]
			\centering  \includegraphics[width=1\linewidth]{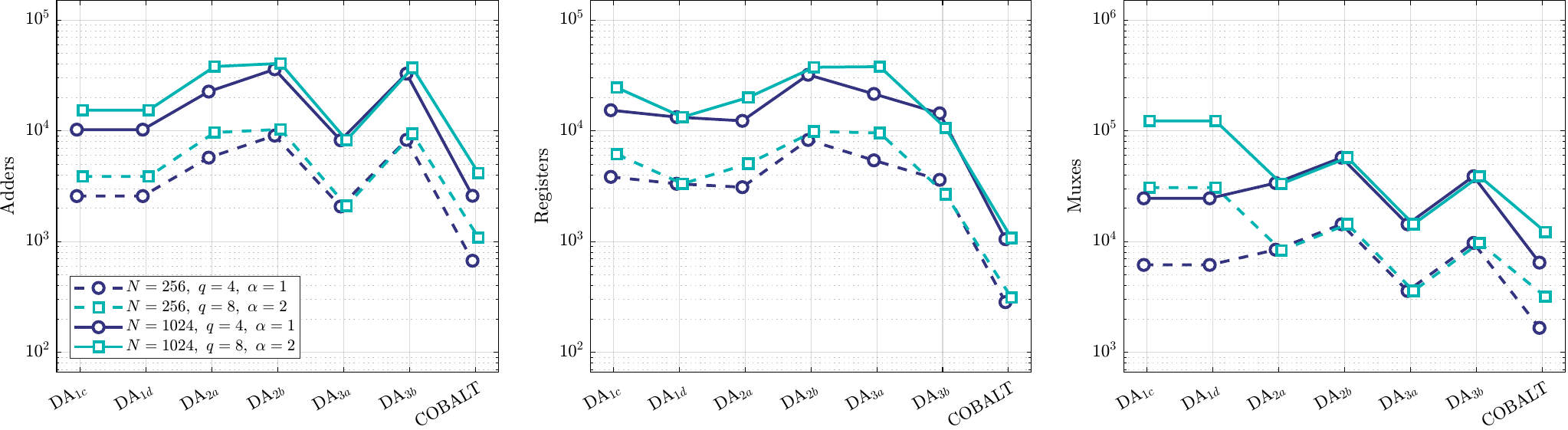}
			\caption{{Illustration of the number of adders, registers, and multiplexers in different DA-based LSTM accelerators with $N=256, 1024$ for $q=4\hspace{0.1cm}(\alpha=1), 8 \hspace{0.1cm}(\alpha=2)$. {Note XOR/XNOR gates are expressed in NAND-equivalents and included with multiplexer count in the same plot (right)}.}}\label{fig14}
		\end{figure*}
		The optimized model was coded in VHDL, functionally verified through simulation, then synthesized,
		placed, and routed in Vivado. {Post-implementation power was estimated using the Xilinx Power Analyzer in
			vector-based (simulation-driven) mode rather than its default vectorless mode: post-implementation timing
			simulation is run with representative input stimuli from the TIMIT dataset~\cite{garofolo1993darpa}, and
			the resulting switching activity is captured in a \texttt{.saif} file and back-annotated into the power
			analysis tool together with the P\&R timing data. This captures the actual toggle rates of nets and
			registers under real data, giving conservative power estimates using a similar methodology to \cite{yan2025accurate}.} Fixed-point adders, multiplexers, and
		registers were implemented using slice LUTs and FFs, with compressed weights, biases, and activations
		held in the distributed-RAM buffers.
		
		The overall system, including on-chip buffers (wRAM, rRAM, bRAM, $\phi$RAM, cRAM, yRAM, BUF$_x$, BUF$_y$), COBALT, and the FSM controller, is shown in Fig.~\ref{fig:system}. The FSM sequences data movement, gate computations, and state updates, manages memory reads/writes, and loads bias terms $\mathbf{b}=[b_1,\dots,b_N]$ via the accumulator unit (AU) to produce
		\begin{gather}
			{\mathbf{v}} =    
			\begin{cases}
				{\mathbf{v}}+{\mathbf{R}}^{T}{\mathbf{t}},  & \text{if}\ {\text{RST}}= 0\\
				{\mathbf{v}}+{\mathbf{b}}, & \text{otherwise}
			\end{cases}
		\end{gather}
		where ${\mathbf{R}}^{T}{\mathbf{t}} = {\mathbf{X}}^T{\mathbf{w}}$ or ${\mathbf{Y}}^{T}{\mathbf{r}}$. The OBC-MVM performs the most resource-intensive computations. Multiple LUTs followed by SA units accelerate the OBC-MVMs in parallel. It processes the circulant input ${\mathbf{X}}^t$ and recurrent output ${\mathbf{Y}}^{t-1}$ with parameters $\mathbf{w}_s$ and $\mathbf{r}_s$ ($s \in \{i, o, f, z\}$). Weights and recurrent weights are stored in wRAM and rRAM, biases in bRAM, and activations—including $\mathbf{x}^t$ and the initial hidden state $\mathbf{y}^0$—are buffered in BUF$_x$ and BUF$_y$. OBC-MVM outputs, after bias addition, pass through nonlinear functions $\phi(\cdot) \in \{\sigma(\cdot), \tanh(\cdot)\}$ implemented via $\phi$RAM to generate gate outputs $\mathbf{i}^t$, $\mathbf{f}^t$, $\mathbf{o}^t$, and $\mathbf{z}^t$. EWMs are applied to $\mathbf{c}^t$ and $\mathbf{y}^t$ as in (2)–(3), implemented as a 1-column OBC-MVM. Partial products ${\pm(\cdot)}/{2}$ are accumulated via SA units to produce EWM outputs. The previous cell state $\mathbf{c}^{t-1}$ is read from cRAM, updated to $\mathbf{c}^t$, and written back, while $\mathbf{y}^{t-1}$ is stored in BUF$_y$ and $\mathbf{y}^t$ in yRAM for the next time step.

		\section{Results and Discussion}
		\subsection{Comparison of Different DA-based LSTM baselines}
		For clarity, the existing DA-based LSTM accelerators~\cite{alhartomi2023low,khan2022architectural,khan2024digit}
		are denoted as DA$_{1a}$--DA$_{1d}$ (four variants), DA$_{2a}$--DA$_{2b}$ (two variants), and
		DA$_{3a}$--DA$_{3b}$ (two variants). All use circulant/block-circulant weight matrices but differ in LUT
		topology (parallel vs.\ serial, TC vs.\ OBC). DA$_1$ adopts full or partial parallel PPs for hardware LUTs,
		while DA$_{2a}$ and DA$_{3a}$ use serial hardware LUTs in OBC and TC form, reducing complexity at the
		expense of higher latency, whereas DA$_{2b}$ and DA$_{3b}$ improve clock rate through high-radix DA, SA
		unfolding, and fine-grained pipelining. In contrast, COBALT reformulates the circulant/block-circulant MVM
		into its input-matrix--weight-vector form, generating PPs directly from the fewer inputs in pairs rather
		than the more numerous weight parameters via column-swapping on the PPs matrix, which avoids the registers
		that circular shifters and tapped-delay units impose on DA$_2$ and DA$_3$. COBALT further avoids the
		fine-grained pipelining stages employed by DA$_{1d}$, DA$_{2b}$, and DA$_{3b}$, which contributes to the
		hardware savings shown in Fig.~\ref{fig14}. For a fair comparison, only the radix-2 variants of DA$_2$/DA$_3$ are considered, along with DA$_{1c}$ and DA$_{1d}$, while DA$_{1a}$ and DA$_{1b}$ are omitted due to their substantial hardware cost.
		\begin{figure}[t]
			\centering  \includegraphics[width=0.78\linewidth]{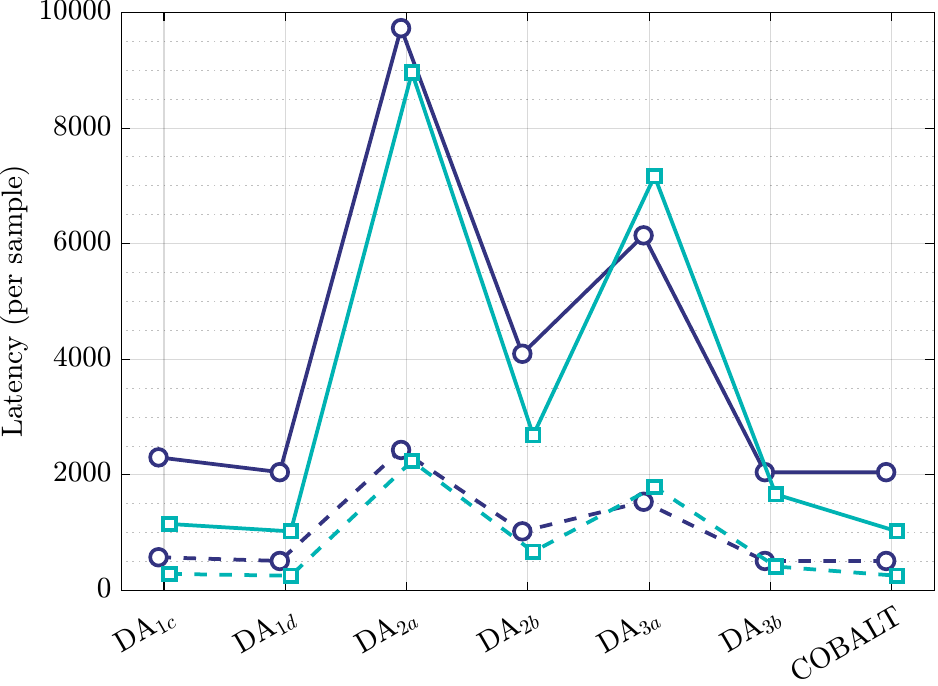}
			\caption{Illustration of the latency involved in different DA-based LSTM accelerators, where the legend used is the same as in Fig. \ref{fig14}.}\label{fig15}
		\end{figure}
		
		In terms of hardware complexity, PPs in DA$_1$--DA$_3$ are generated from unoptimized structures, so
		the LUT size and register cost both grow with the weight matrix dimension, and the gap widens
		continuously as the matrix dimension increases. In COBALT, the sharing of PPGs and the fewer SA units,
		along with correction units required for block-circulant MVM, reduce hardware complexity significantly
		across all three cost categories. At $N=256$ and $q=4$, COBALT requires 67.6\% fewer adders, 94.8\% fewer registers, and 53.6\% fewer multiplexers than DA$_{3a}$. At $N=1024$, these savings increase slightly to 68.5\%, 95.1\%, and 54.9\%, respectively. Against DA$_{1d}$, COBALT achieves savings of 73.9\% in adders, 91.5\% in registers, and 73.0\% in multiplexers at $N=256$, increasing to 74.7\%, 92.1\%, and 73.7\%, respectively, at $N=1024$. As shown in Fig.~\ref{fig14}, the DA$_2$ and DA$_3$ variants generally incur higher hardware costs than the DA$_1$ variants, particularly in adders and multiplexers, due to the additional logic associated with their radix-2 and shift-and-add EWM structures. COBALT remains below the compared designs across all three metrics. The disproportionately
		large register and multiplexer savings against both DA$_{3a}$ and DA$_{1d}$ reflect COBALT's register-free
		PPG/PPS structure, which eliminates the FIFO/tapped-delay units and pipeline-register overhead that
		DA$_{3a}$'s SA unit and DA$_{1d}$'s parallel PP realization both incur. These savings are not fixed
		quantities: as $N$ or $q$ grows, the weight-indexed cost terms in DA$_1$--DA$_3$ scale with $p=N/q$, while
		COBALT's register cost stays near-constant due to block-circulant optimization. Consequently, the register and
		multiplexer advantages widen slightly with $N$, a trend confirmed by the plots, while the adder savings
		remain comparatively flat since both COBALT's and the baselines' adder costs scale similarly with $p$. In terms of latency, DA$_2$ and DA$_3$ incur high delay owing to their serial design, and the additional
		clock cycles required per sub-MVM further worsen the overall latency, as shown in Fig.~\ref{fig15}. All
		designs' latencies scale linearly with $p=N/q$, since sub-MVM optimizations have not been explored for
		DA$_1$--DA$_3$; COBALT's latency is additionally scaled down by the parallelism factor $\alpha=q/4$, so it
		still grows with $N$ but far more slowly than the others, whereas DA$_2$ and DA$_3$ see their unfolded
		latency grow proportionally, with no compensating factor.

		Table~\ref{table2} compares FPGA implementations for $N=256$, $q=4$ on the ZCU104 platform at
		200\,MHz, reporting LUTs, FFs, power, throughput (TP), and energy per sample (EPS). COBALT achieves the lowest
		EPS: 81.2\% better than DA$_{1c}$ and 84.5\% better than DA$_{2b}$. DA$_{1c}$'s parallel PP
		generation and DA$_{2b}$'s fine-grained pipelining both trade additional logic and register switching
		activity for higher clock rate---reducing latency per output but increasing dynamic power and footprint
		simultaneously. COBALT's register-free path reduces both together.

		\begin{table}[t]
			\caption{FPGA results of Different DA-based LSTM accelerators\\ with $N=256$ for $q=4$ at 200 MHz}
			\label{table2}
			\centering
			\resizebox{1\linewidth}{!}{%
				\begin{threeparttable}
					\begin{tabular}{|l||c|c|c|c|c|c|}
						\hline\hline
						{\bf Design} & {\bf DSPs}  & {\bf LUTs} & {\bf FFs}  & {\bf Power} (W) & {\bf TP} (Msps) & {\bf EPS} (nJ/Sa) \\ \hline\hline
						{\bf DA$_{1c}$} \cite{alhartomi2023low} & 12 & 22.38K & 19.98K & 2.72 & 3.125 & 0.870 \\ \hline
						{\bf DA$_{1d}$} \cite{alhartomi2023low} & 12 & 21.32K & 17.21K & 2.55 & 1.56 & 1.635 \\ \hline
						{\bf DA$_{2a}$} \cite{khan2022architectural} & 0 & 35.23K & 26.26K & 3.48 & 0.520 & 6.692 \\ \hline
						{\bf DA$_{2b}$} \cite{khan2022architectural} & 0 & 48.92K & 35.51K & 4.85 & 4.690 & 1.034 \\ \hline
						{\bf DA$_{3a}$} \cite{khan2024digit} & 0 & 28.08K & 19.85K & 3.06 & 0.520 & 5.885 \\ \hline
						{\bf DA$_{3b}$} \cite{khan2024digit}& 0 & 33.64K & 23.88K & 3.98 & 2.605 & 1.528 \\ \hline
						{\bf COBALT}  & 0 & 7.69K & 4.98K & 0.22 & 1.565 & 0.141 \\ \hline
					\end{tabular}
					\begin{tablenotes}
						\item {Throughput: TP  = $(q/pB)f_{\textnormal{clk}}=(16\alpha^2/NB)f_{\textnormal{clk}}$}; EPS (nJ/Sample) = Power / TP.
					\end{tablenotes}
				\end{threeparttable}
			}
		\end{table}

		\subsection{Comparison with State-of-the-Art LSTM Accelerators}
		Several state-of-the-art LSTM accelerators on FPGA were also considered for the comparison, including
		DeltaRNN \cite{gao2018deltarnn}, C-LSTM \cite{wang2018c}, E-RNN \cite{li2019rnn}, BBS \cite{cao2019efficient},
		E-LSTM \cite{wang2019lstm}, Spartus \cite{gao2022spartus}, CVM \cite{li2023fpga}, V-LSTM \cite{kim2023v},
		and ABS \cite{kim2024auto}. Some retain their original names, while others are simplified for clarity. Key
		metrics—including model size, bit precision, PER, FPGA platform, resource utilization, clock frequency,
		TP, and power—are summarized in Table~\ref{table3}. To compare hardware efficiency across heterogeneous
		FPGA platforms, the equivalent number of slices (ENS) metric \cite{liu2019optimized} is used, converting
		DSPs and BRAMs into slice equivalents and adding LUT-derived slices for a unified, family-independent
		measure. {The power figures follow the methodology reported in the respective works. DeltaRNN, C-LSTM,
			E-RNN, Spartus, and V-LSTM report physically measured power, while BBS and ABS report vendor-estimated
			power using Quartus Prime post-fit and Xilinx Power Estimator, respectively; E-LSTM and CVM do not specify
			their methodology. The compared designs also span different process nodes, from 16\,nm (e.g., ABS) to
			28\,nm (e.g., C-LSTM and E-RNN), which can affect cross-node energy-efficiency comparisons
			\cite{gao2018deltarnn,li2019rnn}. These details are provided in the table footnotes for completeness.}

		DeltaRNN exploits temporal sparsity in GRUs to skip redundant computation but is limited in scalability
		due to its small model size. C-LSTM and E-RNN leverage structured (block-circulant) matrices for efficient
		inference; {the DSP/BRAM/LUT/FF figures reported here for C-LSTM are derived from the percentage
			utilization values in \cite{wang2018c} against the ADM-PCIE-7V3's known resource totals, as absolute
			counts are not stated in the source, and should therefore be treated as estimated rather than directly
			reported.} E-LSTM and BBS apply structured pruning to reduce model size and power while maintaining
		accuracy; both are implemented on Intel Arria~10 devices, therefore the DSP/BRAM/LUT/FF figures in
		Table~\ref{table3} are approximate Xilinx-equivalent conversions rather than native ALM/M20K counts, as
		indicated in the table footnotes. Spartus introduces a scalable architecture with hybrid sparsity and
		low-precision computation for high TP. CVM and V-LSTM both prioritize single-sample (batch-1) latency and
		on-chip memory footprint over batched throughput—CVM through a configurable MVM engine for variable-length
		sequences and mixed precision, and V-LSTM through bit-serial processing with flexible precision scaling
		and a Viterbi-based pruning scheme that minimizes DRAM bandwidth. This design emphasis accounts for their
		comparatively low TP figures (0.08 and 0.02\,Msps, respectively) relative to throughput-oriented designs
		such as E-RNN and Spartus, reflecting a different operating point rather than a difference in design
		quality. ABS applies automated block-structured pruning with hardware co-design for balanced accuracy and
		resource usage.
		
		\subsubsection{Resource Utilization}
		In terms of accuracy, COBALT achieves a PER of {20.4--22.2\%}, comparable to the top-performing
		designs in Table~\ref{table3}—Spartus (21.8$\pm$0.3\%) and CVM (21.29\%)—and better than C-LSTM (24.6\%),
		BBS (23.6\%), and V-LSTM (24.11\%). {This range comes from an independent TIMIT hardware-validation run at $N=1024$, separate from the $N=256$ training-strategy sweep described earlier. It uses the same curriculum retraining and FxP8 ADMM/STE quantization strategy at the block-circulant $q=64$ operating point selected for hardware evaluation. The resulting PER is modestly lower range observed at $N=256$, consistent with the expected capacity benefit of the larger hidden layer.} In terms of hardware efficiency, COBALT requires substantially
		fewer LUTs and FFs than the highest-throughput designs: {32.47K} LUTs and {21.57K} FFs versus
		49.9K/39.6K for Spartus and 396.1K/529.7K for ABS. This is most clearly reflected in ENS, where COBALT's
		{8.12K} slices is the lowest among all compared designs—markedly below Spartus (40.63K), C-LSTM
		($\sim$512K, estimated), ABS ($\sim$705K), E-RNN (523.94K), and BBS (1120.41K), the last of which relies
		on 6.8K DSPs and 2.1K BRAMs (Arria~10-equivalent) to reach its throughput. This underscores COBALT's
		suitability for resource-constrained or lower-end FPGA platforms where DSP and BRAM blocks are scarce,
		oversubscribed by other system components, or unavailable altogether.

		\begin{table*}[]
			\caption{Performance Comparison of the Different State-of-the-Art FPGA-based LSTM Accelerators}\label{table3}
			\centering
			\resizebox{1\linewidth}{!}{%
				\begin{threeparttable}
					\begin{tabular}{|l||c|c|c|c|c|c|c|c|c|c|}
						\hline\hline
						{\bf Design} & {\bf DeltaRNN} \cite{gao2018deltarnn} & {\bf C-LSTM}  \cite{wang2018c} & {\bf E-RNN } \cite{li2019rnn} & {\bf BBS } \cite{cao2019efficient} & {\bf E-LSTM } \cite{wang2019lstm} & {\bf Spartus} \cite{gao2022spartus} & {\bf CVM} \cite{li2023fpga} & {\bf V-LSTM} \cite{kim2023v} & {\bf ABS} \cite{kim2024auto} & {\bf COBALT} \\ \hline\hline
						
						{\bf Model size (\textit{N})} & 128 & 512 & 1024 & 1024 & 1024 & 1024 & 1024 & 1024 & 1024 & 1024 {($q=64$)} \\ \hline
						{\bf Precision} & INT16/16 & INT16/16 & INT16/16 & INT16/16 & INT8/8 & INT16/8 & INT12/12 & INT16/4 & INT16/16 & INT8/8 \\ \hline
						{\bf PER (\%)} & -- & 24.6 & {20.3\tnote{f}} & 23.6 & 23.2 & 21.8$\pm$0.3 & 21.29 & 24.11 & -- & {20.3-22.2} \\ \hline
						{\bf FPGA Platform} & XC7Z100 & 7V3 & XC/VX690T & GX1150 & SX660 & XC7Z100 & ZCU102 & VC709 & XCKU115 & ZCU104 \\ \hline
						{\bf DSP} & 768 & 2.8K{\tnote{d}} & 2.9K & 6.8K{\tnote{e}} & 10{\tnote{e}} & 65 & 1457 & 24 & 3.9K & 0 \\ \hline
						{\bf BRAM} & 457.5 & 931{\tnote{d}} & 1.4K & 2.1K{\tnote{e}} & 828{\tnote{e}} & 185 & 610 & 97 & 1.77K & 0 \\ \hline
						{\bf LUT} & 261.4K & 475.2K{\tnote{d}} & 257.2K & 724.3K{\tnote{e}} & 490.3K{\tnote{e}} & 49.9K & 187.1K & 350.5K & 396.1K & {32.47K} \\ \hline
						{\bf FF} & 119.3K & 206.6K{\tnote{d}} & 479.2K & --{\tnote{e}} & 173.3K{\tnote{e}} & 39.6K & 289.8K & 172.4K & 529.7K & {21.57K} \\ \hline
						{\bf Freq. (MHz)} & 125 & 200 & 200 & 200 & 200 & 200 & 200 & 200 & 150 & {200} \\ \hline
						{\bf TP (Msps)} & 192 & 51.2 & {1638.4} & {51.2} & 282.2 & 204.8 & 0.08 & 0.02 & 307.2 & {100$^{*}$} \\ \hline
						{\bf Power (W)} & {5.5}{\tnote{a}} & 23{\tnote{b}} & 25{\tnote{b}} & 19.1{\tnote{c}} & 15.9 & 8.4 & 15.0 & 20.2 & 9.674{\tnote{c}} & {1.33} \\ \hline
						{\bf EPS (nJ/Sample)}{$^{**}$} & {28.65} & {449.22} & {15.26} & {373.05} & {56.34} & {41.0} & {181180} & {961910} & {31.49} & {13.3} \\ \hline
						{{\bf ENS}} & {{197.14K}} & {{512.28K}}{\tnote{d}} & {523.94K} & {1120.41K} & {219.81K} & 40.63K & {{266.85K}} & {{101.35K}} & {{705.36K}} & {8.12K} \\ \hline
						{{\bf TP/ENS (Msps/K)}} & {0.97} & {0.10} & {3.13} & {0.05} & {1.28} & {5.04} & {0.0003} & {0.0002} & {0.44} & {12.32} \\ \hline
						{{\bf EPE (TP/ENS $\div$ PER)}} & -- & {0.0041} & {0.1541} & {0.0019} & {0.0553} & {0.2282} & {$1.4 \times 10^{-5}$} & {$8.3\times 10^{-6}$} & -- & {0.56} \\ \hline
						{{\bf EPE (norm.\ to Spartus)}} & -- & {0.02$\times$} & {0.68$\times$} & {0.01$\times$} & {0.24$\times$} & {1.00$\times$} & {$\sim$0$\times$} & {$\sim$0$\times$} & -- & {2.43$\times$} \\ \hline\hline
					\end{tabular}
					\begin{tablenotes}
						\item {DSPs and BRAMs are converted into ENS and add LUT-derived slices. Assumptions are based on \cite{liu2019optimized}: 1 DSP = 102.4 slices, 1 BRAM (assume 18K) = 116.2 slices, 1 slice = 4 LUTs $\Rightarrow$ slices from LUT = LUT / 4, ENS = (LUT / 4) + DSP $\times$ 102.4 + BRAM $\times$ 116.2.} DeltaRNN is a variant of RNNs, Spartus exploits spatio-temporal sparsity. {$^{*}$, $^{**}$: Calculated based on the TP and EPS formulae listed in the footnote of Table \ref{table2}, respectively. EPE (efficiency-per-error) = (TP/ENS)/PER$_{\text{worst}}$, normalizing hardware efficiency by worst-case reported accuracy; designs without reported PER (DeltaRNN, ABS) are excluded. EPE (norm.) is relative to the best prior baseline (Spartus). \tnote{a}DeltaRNN's on-chip FPGA power (5.5\,W) is cross-validated against the Xilinx Power Analyzer estimate (5.503\,W) in \cite{gao2018deltarnn}; board-level power is 7.3\,W.
							\tnote{b} Power for C-LSTM and E-RNN is physically measured on the ADM-PCIE-7V3 platform using TI Fusion
							Power monitoring devices; no power figure is available for the KU060 platform. \tnote{c} Power for BBS and ABS is estimated via vendor tools (Quartus Prime post-fit analysis and
							Xilinx Power Estimator, respectively) rather than measured on hardware. \tnote{d} C-LSTM's \cite{wang2018c} reports only percentage utilization, not absolute DSP/BRAM/LUT/FF counts; values shown are derived from these percentages against the ADM-PCIE-7V3's known resource totals (DSP=3600, BRAM=1470, LUT=859.2K, FF=429.6K), using the FFT16 (block size 16), 7V3 column, which matches the reported 23\,W power. \tnote{e} BBS and E-LSTM are implemented on Intel Arria~10 (ALM/M20K architecture), not Xilinx LUT/FF/BRAM; reported figures are approximate Xilinx-equivalent conversions. \tnote{f} E-RNN's PER is not reported in \cite{li2019rnn}; the 20.3\% figure is Spartus's own reproduction \cite{gao2022spartus}, not the original E-RNN.}
					\end{tablenotes}
				\end{threeparttable}
			}
		\end{table*}
		
		\subsubsection{Power and Energy Efficiency Trade-offs}
		Table~\ref{table3} shows that COBALT has the lowest resource footprint, although it does not achieve the
		highest throughput. COBALT operates at 200\,MHz, reaching 100\,Msps at {1.33 W}, using zero DSPs and
		zero BRAMs, with {32.47K} LUTs and {21.57K} FFs. Despite its lower throughput relative to Spartus
		(100 vs.\ 204.8\,Msps), COBALT achieves a substantially lower energy per sample: {13.3}\,nJ/sample
		versus 41.0\,nJ/sample for Spartus, a {$3.1\times$} improvement, driven by an approximately
		{$6.3\times$} lower power draw ({1.33}\,W vs.\ 8.4\,W) that more than offsets its lower
		throughput. This trade-off--sacrificing raw throughput for a disproportionately larger reduction in
		power--is enabled by COBALT's DSP/BRAM-free, LUT-only architecture. ABS and E-RNN similarly achieve
		higher throughput than COBALT, reaching 307.2\,Msps at 9.674\,W and 1638.4\,Msps at 25\,W, respectively.
		However, both rely heavily on dedicated DSP and memory resources. Thus, COBALT targets a different point
		in the throughput--resource trade-off. Designs such as Spartus, ABS, and E-RNN are preferable when
		throughput is the primary constraint and DSP/BRAM resources are available, while COBALT is attractive for
		resource-constrained or lower-end FPGA deployments where low absolute power and DSP/BRAM-free operation
		are important.
		
		{Raw efficiency metrics such as EPS do not account for accuracy, so a design could appear efficient
			simply by tolerating a higher PER. TP/ENS (throughput per unit hardware cost) and EPE (TP/ENS normalized
			by worst-case PER) address this: COBALT reaches 12.32\,Msps/K-ENS, over $2.4\times$ Spartus's
			5.04\,Msps/K-ENS and well above E-RNN (3.13) and E-LSTM (1.28). Once accuracy is accounted for through EPE, COBALT still leads with 0.56, representing a $2.43\times$ improvement over Spartus, while maintaining comparable accuracy.}

		\section{Conclusion}
		This paper presented COBALT, a bit-serial LSTM accelerator that reformulates circulant MVM as a
		matrix (input)--vector (weight) product under offset-binary coding, and demonstrated how bit-level
		redundancy across output rows can be systematically exposed and removed. The column-swapping
		scheme and paired-row correction unit together cut the number of PP generators, selectors, and
		adders required for both circulant and block-circulant MVM, while relocation of the shift-accumulate
		unit yields further savings in the block-circulant case. On FPGA, these choices {exceed} the
		efficiency of the state-of-the-art accelerator in~\cite{gao2022spartus}. {As MVM is a core,
			increasingly dominant operation across edge AI workloads beyond LSTMs, COBALT's redundancy-exploitation
			principle offers a structure-agnostic building block for future resource-constrained accelerators.}

		\bibliographystyle{IEEEtran}
		\bibliography{main}
		
		\vspace{-1.5cm}
		
		\begin{IEEEbiography}
			[{\includegraphics[width=1.07in,height=1.28in,clip,keepaspectratio]{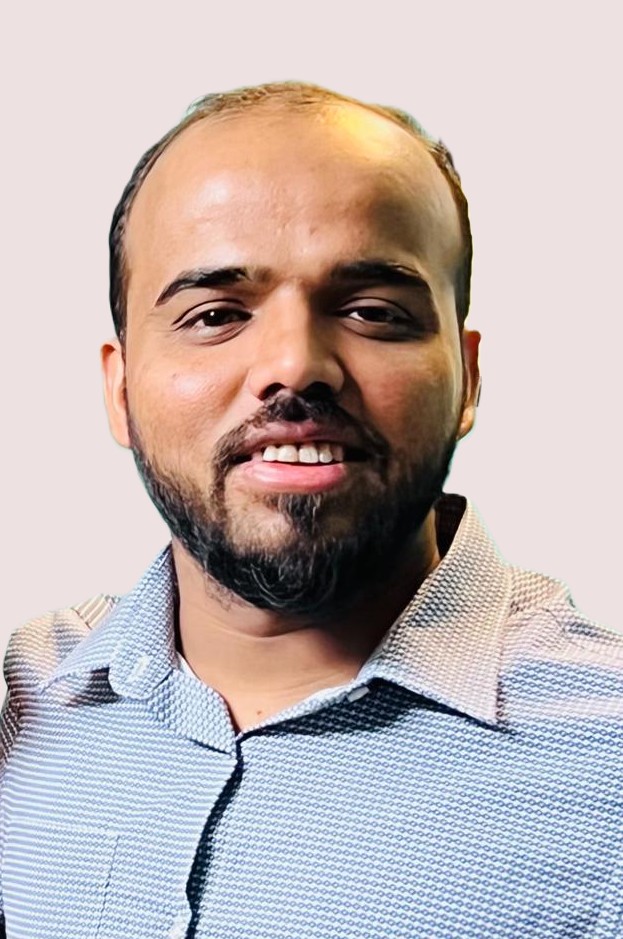}}]
			{Mohd. Tasleem Khan} (Senior Member, IEEE) received his B.Tech in Electronics from AMU, India, in 2013, and his Ph.D. in VLSI from IIT Guwahati, India, in 2019. He was a Principal Engineer at TSMC, Taiwan, and an Assistant Professor at IIT Dhanbad, India. From 2021 to 2024, he was a Postdoctoral Research Associate at Linköping University, Sweden. He is currently an Assistant Professor at Heriot-Watt University, Edinburgh, UK. His research focuses on algorithms and architectures for VLSI implementation in Machine Learning/AI, Signal Processing, and Communication Systems. He serves as Associate Editor for IEEE Signal Processing Letters, IEEE TNNLS, and IEEE TASE, is on the Editorial Board of IEEE Embedded Systems Letters, is currently a Guest Editor for IEEE Transactions on Circuits and Systems II, and reviews for IEEE TCAS-I, II, TVLSI, and related journals. He is a member of the IEEE Signal Processing and Circuits and Systems Societies, and has served as Area Chair for IEEE ICASSP’26, IEEE ICJNN’25, and TPC member for ISVLSI’25, ICDCS’25, and SaTC’25.
			
		\end{IEEEbiography}
		%\vspace{-1.2cm}
		
	\end{document}